%% file: bulk_inversion.tex
\RequirePackage{pdf14}

\pdfoutput=1
\documentclass[letterpaper]{article}
\usepackage[usenames,dvipsnames,table]{xcolor}
\usepackage{graphicx}
\usepackage{caption}
\usepackage{subcaption}
\usepackage{array,setspace,mathrsfs,amsthm,amsfonts,amsmath,colonequals,mathtools}
\usepackage{jheppub}
\usepackage{afterpage}
\usepackage{xspace}
\usepackage[normalem]{ulem}
\usepackage{cleveref}
\usepackage{tikz}

\input{topmatter}

\title{A Lorentzian inversion formula for defect CFT}
\author{Pedro Liendo,}
\author{Yannick Linke,}
\author{Volker Schomerus.}

\affiliation{DESY Hamburg, Theory Group, Notkestra{\ss}e 85, D-22607 Hamburg, Germany}

\emailAdd{pedro.liendo@desy.de}
\emailAdd{yannick.linke@desy.de}
\emailAdd{volker.schomerus@desy.de}

\preprint{DESY 19-039}

\bigskip
\abstract{
	We present a Lorentzian inversion formula valid for any defect CFT that extracts the bulk channel CFT data as an analytic function of the spin variable. This result complements the already obtained inversion formula for the corresponding defect channel, and makes it now possible to implement the analytic bootstrap program for defect CFT, by going back and forth between both channels. A crucial role in our derivation is played by the Calogero-Sutherland description of defect blocks which we review. As first applications we obtain the large-spin limit of bulk CFT data necessary to reproduce the defect identity, and also calculate the bulk data of the twist defect of the $3d$ Ising model to first order in the $\epsilon$-expansion.
}


\begin{document}
\setcounter{tocdepth}{2}
\maketitle
\setcounter{page}{1}

\input{Intro}

%
\input{review}

%
\input{inversion}
%
\input{defect_identity}
%
\input{3dIsing}

%
\input{conclusions}

\appendix
\input{AppendixA}

\bibliography{bulk_inversion}
\bibliographystyle{JHEP}

\end{document}

%% file: topmatter.tex
\let\oldbfseries=\bfseries
\let\oldmdseries=\mdseries
\let\oldnormalfont=\normalfont
\renewcommand{\bfseries}{\oldbfseries\boldmath}
\renewcommand{\mdseries}{\oldmdseries\unboldmath}
\renewcommand{\normalfont}{\oldnormalfont\unboldmath}

\makeatletter
\newlength{\apb@width}
\newcommand{\autoparbox}[2][c]{\settowidth{\apb@width}{#2}\parbox[#1]{\apb@width}{#2}}

\makeatother

\newcommand{\beqa}{\begin{eqnarray}}
\newcommand{\eeqa}{\end{eqnarray}}
\newcommand{\beq}{\begin{equation}}
\newcommand{\eeq}{\end{equation}}

\newcommand{\D}{\Delta}

\newcommand{\op}[1]{\operatorname{#1}}
\newcommand{\ie}{i.\,e.\ }
\newcommand{\eg}{e.\,g.\ }

\newmuskip\pFqskip
\mathchardef\pFcomma=\mathcode`,

\newcommand*\pFq[5]{%
   \begingroup
   \begingroup\lccode`~=`,
     \lowercase{\endgroup\def~}{\pFcomma\mkern\pFqskip}%
   \mathcode`,=\string"8000
   {}_{#1}F_{#2}\biggl(\genfrac..{0pt}{}{#3}{#4};#5\biggr)%
   \endgroup
}

\newcommand{\cblock}[4]{
	\mathop{#1}\left(\begin{matrix}
		#2 \\
		#3
	\end{matrix}
	; #4 \right)
}

%% file: Intro.tex
%!TEX root = ../bulk_inversion.tex
%%%%%%%%%%%%%%%%%%%%%%%%%%%%%%%%%%%%%%%%%%%%%

\section{Introduction}

Non-local operators and defects are important observables in field theory both from a theoretical and phenomenological point of view. Typical examples in gauge theories are Wilson and 't Hooft lines, while boundaries and interfaces are common in condensed matter systems. In the context of conformal field theories (CFTs), extended objects break a portion of the conformal symmetry to a subgroup, which is nevertheless powerful enough for the application of the conformal bootstrap program. The main motivation behind this work is the generalization of analytic bootstrap techniques to the case of defect CFTs.

A thorough analysis of defect CFTs in higher dimensions was initiated in \cite{Billo:2016cpy}.
The basic quantities that characterize a defect are the
one-point functions, \ie the one-point correlator of a local bulk operator $\mathcal{O}$
in the presence of a defect, and the two-point functions between a bulk field $\mathcal{O}$ and a defect field  $\mathcal{\widehat{O}}$  \ie
fields that can be inserted along the defect.
These two types of correlators are fixed by conformal symmetry
up to an overall constant: $\alpha_\mathcal{O}$ for the one-point correlator and
$b_{\mathcal{O} \widehat{\mathcal{O}}}$ for the bulk-to-defect correlator, and are therefore
reminiscent of three-point functions of bulk fields which are also fixed by symmetry.

The first defect correlator that cannot be fixed by kinematics is the two-point
function of bulk fields.
The geometric data of a conformal defect with $p<d-1$ and two bulk insertion
points can be characterized by two conformal invariants. Hence, up to a simple prefactor, defect
two-point functions depend on two variables, much as the four-point function of bulk
fields depends on two cross-ratios.

A defect two-point function can be calculated in two different ways, either by
applying the bulk operator expansion to the two bulk fields and subsequent evaluation
of the defect one-point functions, or by a bulk-defect operator product expansion (OPE) which
rewrites one of the bulk field insertions as an infinite sum over defect fields, and subsequent evaluation of the
resulting bulk-to-defect correlator.
These two computational schemes are referred to as \textit{bulk channel} and \textit{defect channel} respectively, and are captured by a conformal block expansion \cite{Billo:2016cpy,Lauria:2017wav,Isachenkov:2018pef,Lauria:2018klo}.
They resemble the $s-$ and $t-$channels of bulk four-point functions, but while the latter are
very much of the same form, \ie both channels involve the same type of
conformal blocks and product of bulk OPE coefficients, the bulk and
defect channel of defect two-point functions possess a very different structure,
with different conformal blocks and coefficients. In
particular, even for two identical bulk fields, the coefficients in the defect
channel are positive while those in the bulk channel may not be.

Equality of the two computational schemes we have described gives a consistency condition and is the analog of
crossing symmetry for this correlator.\footnote{For defect two-point functions the terms ``cross-ratio'' and ``crossing symmetry'' are not quite accurate, however they have become standard in the defect literature and so will use them.}
This consistency condition is very powerful but difficult to analyse. The numerical bootstrap of \cite{Rattazzi:2008pe}
relies heavily on the positivity of coefficients in the
$s-$ and $t-$channel, so the lack of positivity in the bulk channel of the defect
two-point functions presents a significant obstacle.
As a side comment, let us point out however that if one considers operators constrained to the defect, positivity is restored and numerical bootstrap techniques can be used. This setup has gotten some attention recently in particular in the context of $1d$ CFTs \cite{Gaiotto:2013nva, Hogervorst:2017sfd, Mazac:2016qev, Giombi:2017cqn, Liendo:2018ukf,Mazac:2018ycv, Mazac:2018qmi, Kiryu:2018phb, Arkani-Hamed:2018ign, Beccaria:2019dws}, which have a natural interpretation in terms of the theory living on a line defect.

Analytical bootstrap methods on the other hand do not require positivity and are suitable for the study of defect two-point functions.
While the sheer number of bulk primaries, \ie terms on the
conformal block decompositions for the two channels, has so far impeded full analytical
solutions in dimension $d>2$, there exists a lightcone limit in which the
complexity of the equations can be reduced \cite{Fitzpatrick:2014vua,Komargodski:2012ek}.
This program goes by the name of the lightcone bootstrap, it allows for systematic twist expansions and has provided a wealth of interesting results on the dynamics probed by fields of large spin.

A powerful way to study the lightcone limit for bulk four-point functions is the Lorentzian
inversion formula derived in \cite{Caron-Huot:2017vep} (see also \cite{Simmons-Duffin:2017nub} for an alternative derivation). Caron-Huot's formula
recovers the dynamical information in the various channels of a four-point function
by performing a certain integral over the space of cross-ratios
in Lorentzian kinematics of which the lightcone limit is a particular corner. The
original Lorentzian inversion formula has been extended in several different ways, \eg
to four-point functions of fields with spin in \cite{Kravchuk:2018htv} or thermal correlators \cite{Iliesiu:2018fao}.
Most relevant for our discussion is the defect channel inversion formula of defect
two-point functions proposed in \cite{Lemos:2017vnx}. This formula allows
to recover dynamical information in the defect channel, such as \eg the conformal
weights of defect fields, from a certain integral over cross-ratios in the Lorentzian
regime.

Since the two channels of a defect two-point functions are different, a complete
implementation of the lightcone bootstrap for defect CFT necessitates the complementary bulk channel inversion formula.
Caron-Huot's original derivation
as well as the subsequent extensions, require solid knowledge of conformal blocks for
the involved channel. While blocks for the defect channel of defect two-point
functions were already known in closed-form \cite{Billo:2016cpy}, a systematic theory of bulk channel blocks
was developed recently in \cite{Isachenkov:2018pef}, by relying on the connection between conformal blocks and Calogero-Sutherland models \cite{Isachenkov:2016gim}.

Armed with a complete theory of blocks for the bulk channel of defect two-point
functions we are now able to propose the associated Lorentzian inversion formula.
This formula, which is the central result of our work, is stated in equation
\eqref{eq:defInversion} below. After presenting our derivation we shall consider
two immediate applications. First, we derive the leading large spin
behaviour of the bulk-to-defect couplings $\alpha_{\mathcal{O}}$ for certain
families of bulk fields in a large class of defect CFTs.
A second application concerns a specific defect theory, namely the twist
defect of the $3d$ Ising model at the critical point \cite{Billo:2013jda,Gaiotto:2013nva}. In this case we compute
the one-point functions appearing in the $\sigma \times \sigma$ OPE to leading order in $\epsilon = 4-d$.

Let us now outline the rest of this work. In the next section we review some
relevant previous results. In particular we recall how to construct the conformal
blocks for both the bulk and the defect channel. In addition, we shall discuss
the existing inversion formulas for defect two-point function, namely the
Lorentzian inversion formula for the defect channel \cite{Lemos:2017vnx}, and
the Euclidean inversion formula for the bulk channel \cite{Isachenkov:2018pef}.
The latter may be regarded as an ancestor of the Lorentzian inversion formula
we are going to state at the beginning of section \ref{sec:lorentzianInversion}. The rest of section \ref{sec:lorentzianInversion} is
then devoted to a derivation of this formula. After a lightning review of the
Calogero-Sutherland approach to defect blocks from \cite{Isachenkov:2018pef}, we
present convincing evidence for our Lorentzian inversion formula for the bulk
channel of defect two-point functions. Section \ref{sec:lcDef} and \ref{sec:3dIsing_twist_defect} are devoted to
applications. The first application in section \ref{sec:lcDef} is prototypical for the use
of Lorentzian inversion formulas, and it will result in expressions for the
leading large-spin behavior of one-point functions $\alpha_\mathcal{O}$.
It applies under certain conditions on the defect which will be stated and
discussed in much detail. Section \ref{sec:3dIsing_twist_defect} contains our application to the twist
defect of the $3d$ Ising model. Our analysis relies on the fact that we are
able to resum a result in \cite{Gaiotto:2013nva} for the first order term in
the $\epsilon = 4-d$ expansion of the two-point function with
two spin fields. Our analytic expression for this quantity enables us to
compute the bulk-to-defect couplings $\alpha_\mathcal{O}$ for all bulk primaries
$\mathcal{O}$ that appear in the OPE of the bulk spin
field $\sigma$. 
The paper concludes with a summary and outlook to further
directions.

%% file: review.tex
%!TEX root = ../bulk_inversion.tex
%%%%%%%%%%%%%%%%%%%%%%%%%%%%%%%%%%%%%%%%%%%%%

\section{Preliminaries}
\label{sec:preliminaries}

In this section we provide most of the background material that is necessary to formulate the problem
and to perform our subsequent analysis. In the first subsection we introduce defect two-point functions and discuss
their conformal block decomposition in both the bulk and defect channels.
In the second subsection we review some basic features of the lightcone bootstrap \cite{Fitzpatrick:2012yx,Komargodski:2012ek,Alday:2016njk,Caron-Huot:2017vep} adapted to the defect setup, and discuss some relevant inversion formulas. Euclidean inversion formulas are known for the defect \cite{Lemos:2017vnx} and  bulk \cite{Isachenkov:2018pef} channel, but so far a Lorentzian version was only obtained for the defect case; see \cite{Lemos:2017vnx} and our review in subsection \ref{sec:lcBootstrap}.

\subsection{Two-point functions in defect CFTs}
\label{sec:2ptDefectCFTs}

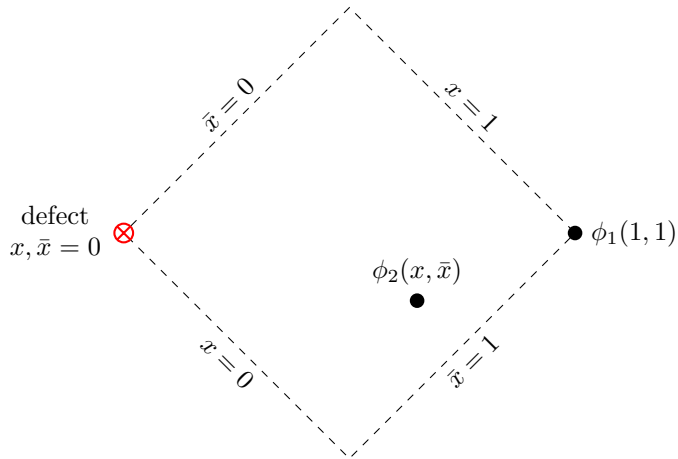
\begin{figure}
	\centering
	\mbox{\begin{tikzpicture}[scale = 3, cross/.style={path picture={
				\draw[red]
				(path picture bounding box.south east) -- (path picture bounding box.north west) (path picture bounding box.south west) -- (path picture bounding box.north east);
		}}]	
		\node[draw, red, circle, cross, scale = .75, thick, label={[text width=1.3cm, align=center]left:defect \\ $x,\bar{x}=0$}] (defect) at (-1,0) {};
		\node[draw, circle, fill, scale = .5, thick, label={right:$\phi_1(1,1)$}] (phi1) at (1,0) {};
		\node[draw, circle, fill, scale = .5, thick, label={above:$\phi_2(x,\bar{x})$}] (phi2) at (.3,-.3) {};
		
		\draw[dashed] (defect) -- node[above, sloped] {$\bar{x}=0$} (0,1) -- node[above, sloped] {$x=1$} (phi1) -- node[below, sloped] {$\bar{x}=1$} (0,-1) -- node[below, sloped] {$x=0$} (defect);
		\end{tikzpicture}}
	\caption{Two-point function configuration in a plane orthogonal to the defect.
	The flat defect is at the origin while the operators $\phi_1$ and $\phi_2$ are at points
	$(1,1)$ and $(x,\bar{x})$, respectively.}
	\label{fig:configuration}
\end{figure}

Let us consider a flat or spherical defect $\mathcal{D}^{(p)}$ of dimension $p$ in $d$ spacetime
dimensions. In this work the defect will always be space-like. In what follows we are mainly
interested in the two-point function of bulk scalar primaries $\phi_1$ and $\phi_2$ in the
presence of the defect. This configuration depends on two cross-ratios which we denote by $x$
and $\bar x$. We refer to \cite{Billo:2016cpy} for a general introduction to the topic of
defect CFTs. If we place the $p$-dimensional defect along $x_i= 0, i=1, \dots, p$, the
two-point function reads
\begin{equation}\label{eq:2ptFct}
\langle \mathcal{D}^{(p)} \phi_1(x_1) \phi_2(x_2) \rangle = \frac{\mathcal{F}(x,\bar x )}
{|x_1^\perp|^{\Delta_1}|x_2^\perp|^{\Delta_2}}\ .
\end{equation}
Here $|x_i^\perp|$ denotes the transverse distance of the insertion points $x_i$ from
the defect and the cross-ratios are given by
\begin{equation}\label{eq:xxbar}
\frac{x+\bar x}{2(x\bar x)^{1/2}} = \frac{x_1^\perp \cdot x_2^\perp}{|x_1^\perp|
	|x_2^\perp|} \,, \quad
\frac{(1-x)(1-\bar x)}{(x\bar x)^{1/2}} = \frac{(x_1-x_2)^2}{|x_1^\perp| |x_2^\perp|} \,.
\end{equation}
The geometry is shown in figure \ref{fig:configuration}. In Euclidean signature the cross
ratios $x$ and $\bar x$ are complex conjugate to each other.

The defect two-point function can be expanded in two channels depending on whether we perform an OPE of the two bulk fields, or we expand a bulk field in terms
of defect operators. We denote these two channels as the \textit{bulk} and \textit{defect channel},
respectively. Let us first look at the bulk channel. In this case the associated conformal block
expansion reads
\begin{equation}\label{eq:2ptbulkCBD}
	\mathcal{F}(x, \bar{x}) = \left(\frac{(1-x)(1-\bar{x})}
 {(x \bar{x})^{\frac{1}{2}}} \right)^{-\frac{\Delta_1+\Delta_2}{2}}
	\sum_{ \mathcal{O}} c_{12 \mathcal{O}}a_{\mathcal{O}} f_{\Delta,\ell}(x, \bar x) \,.
\end{equation}
The prefactor on the right-hand-side contains terms that depend on the sum $\Delta_1 +\Delta_2$
of the conformal weights of the two scalar bulk fields. The bulk channel blocks
$$f_{\Delta,\ell}(x, \bar x) \equiv \cblock{f}{p,a,d}{\Delta,\ell}{x,\bar{x}} $$
depend on the external weights only through the difference $2a = \Delta_2-\Delta_1$. In addition,
they possess a very non-trivial dependence on the dimension $d$ of spacetime and the dimension $p$
of the defect, see below. The summation is performed over bulk fields of conformal weight $\Delta$
and spin $\ell$, just as in the more  familiar case of conformal block expansions for four-point
functions of bulk fields. Note however that the coefficients contain different dynamical information. They
are given by the product of a three-point coupling $c_{12 \mathcal{O}}$ and the one-point coupling
$a_{\mathcal{O}}$ of a bulk field in the presence of the defect.

The bulk channel blocks for defect two-point functions of two scalar fields were determined
in \cite{Lauria:2017wav,Isachenkov:2018pef}, see also \cite{Lauria:2018klo} for a generalization to spinning blocks. In this work, the bulk channel blocks was constructed as
a linear combination of two \emph{Harish-Chandra functions} (also called \emph{pure functions}
in \cite{Caron-Huot:2017vep}). To be concrete, let us spell out the following series expansion
\cite{Isachenkov:2017qgn}
\begin{equation}\label{eq:HSfct}
	\begin{aligned}
	f^{HS}_{\Delta,\ell}(x, \bar{x})  = & (x\bar{x})^{\frac{a}{2}}
\sum_{m=0}^{\infty}\sum_{n=0}^{\infty} h_n(\Delta,\ell) h_m(1-\ell,1-\Delta) \frac{\left(\frac{\Delta+\ell}{2}-a\right)_{n-m}}{\left(\frac{\Delta+\ell}{2}-\frac12\right)_{n-m}}
\frac{4^{m-n}}{n!m!}
\\[2mm]
&\qquad
\times
	\pFq{4}{3}{-n,-m,\frac12-a,\frac{\Delta-\ell}{2}-\frac{d}{2}-a+1}{-\frac{\Delta+\ell}{2}-a+1-n,
\frac{\Delta+\ell}{2}-a-m,\frac{\Delta-\ell}{2}-\frac{d}{2}+\frac32}{1} \, (1-x\bar{x})^{\ell-2m}
\\[2mm]
	&\qquad \times \left[(1-x)(1-\bar{x})\right]^{\frac{\Delta-\ell}{2}+m+n} \,
	\pFq{2}{1}{\frac{\Delta+\ell}{2}+a-m+n,\frac{\Delta+\ell}{2}-m+n}
{\Delta+\ell-2m+2n}{1-x\bar{x}} \,,
	\end{aligned}
\end{equation}
with
\begin{equation}
	h_n(\Delta,\ell) = \frac{(\frac{\Delta}{2}-\frac12,\frac{\Delta}{2}-\frac{p}2,\frac{\Delta+\ell}{2}+a)_n}
{(\Delta-\frac{d}{2}+1,\frac{\Delta+\ell}{2}+\frac12)_n} \,.
\end{equation}
Here  $(x)_n = \Gamma(x+n)/\Gamma(x)$ is usual the Pochhammer symbol. The Harish-Chandra functions exhibit
a pure power law behavior in the limit $0 \ll 1-x \ll 1-\bar{x} \ll 1$:
\begin{equation}\label{eq:defblocknormxxb}
	f^\textit{HS}_{\Delta,\ell}(x, \bar{x}) =
	(1-x)^{\frac{\Delta-\ell}{2}}(1-\bar{x})^{\frac{\Delta+\ell}{2}} \times
(1+\text{integer powers of $(1-x)/(1-\bar{x})$, $1-\bar{x}$}) \,.
\end{equation}
We stress that the series expansion in equation \eqref{eq:HSfct} is analytic in spin, a fact that shall become useful
later in the context of the Lorentzian inversion formula. Note that a somewhat similar series expansion for
four-point blocks that was derived in \cite{Dolan:2011dv} does not have this property, and in fact does only
give correct results for integer values of the spin.

Armed with some explicit series expansion for $f^\textit{HS}_{\Delta,\ell}$ we can now use these functions
to construct the bulk channel conformal blocks as
\begin{equation}\label{eq:fblock}
	\begin{aligned}
		f_{\Delta,\ell}(x, \bar{x}) &= f^\textit{HS}_{\Delta,\ell}(x, \bar{x}) \\
		&\qquad+
		\frac{ \Gamma\left( \ell + d - 2 \right)\Gamma\left( -\ell-\frac{d-2}{2} \right) }{ \Gamma\left( \ell + \frac{d-2}{2} \right)\Gamma\left( -\ell \right) }
		\frac{ \Gamma\left( \frac{\ell}{2}+\frac{d-p}{2}-\frac{1}{2} \right)\Gamma\left( -\frac{\ell}{2} + \frac12 \right) }{ \Gamma\left( \frac{\ell}{2}+\frac{d}{2}-\frac{1}{2} \right)\Gamma\left( -\frac{\ell}{2}-\frac{p}{2}+\frac{1}{2} \right) }
		f^{HS}_{\Delta,2-d-\ell}(x, \bar{x})\,.
	\end{aligned}
\end{equation}
Before turning to the defect channel, we note that our formulas for the bulk channel bear some similarities
with corresponding ones in the analysis of bulk four-point functions. This is not surprising since the
defect two-point function in equation \eqref{eq:2ptFct} reduces to the four-point function of scalar bulk fields
when the defect becomes a point-like defect, \ie when $p=0$. In some cases the bulk channel blocks can indeed
be identified with scalar four-point blocks, but this is not true in general. We will review the precise
relation between bulk channel and four-point blocks in subsection \ref{sec:CSapproach} below and also discuss
the differences between these quantities.
\medskip

In the defect channel the conformal block expansion reads
\begin{equation}\label{eq:defectCBD}
	\mathcal{F}(x, \bar{x})
	= \sum_{\widehat{\mathcal{O}}} b_{1  \widehat{\mathcal{O}}} b_{2 \widehat{\mathcal{O}}}
\hat{f}_{\widehat{\Delta},s}(x, \bar{x})
	\,.
\end{equation}
Here, the conformal blocks $\hat{f}_{\widehat{\Delta},s}(x, \bar{x})$ are labeled by the conformal
dimension $\widehat{\Delta}$ of the defect operator $\widehat{\mathcal{O}}$ and by the transverse
spin variable $s$ which is associated to rotations around the defect. General defect fields can
also carry a spin label with respect to the rotation group $SO(p)$ of the defect, but such defect
fields cannot be excited in the defect expansion of scalar bulk fields \cite{Billo:2016cpy}. The coefficients
$b_{k \widehat{\mathcal{O}}}$  are bulk-to-defect couplings and they describe new dynamical data.
The symmetry group of the defect channel factorizes to $SO(1,p+1) \times SO(d-p)$,
and this allows to write the conformal blocks as a product of two
single-variable hypergeometric functions \cite{Billo:2016cpy}:
\begin{equation}\label{eq:defBlock}
	\hat{f}_{\widehat{\Delta},s}(x, \bar{x}) = x^{\frac{\widehat{\Delta}-s}{2}}
\bar{x}^{\frac{\widehat{\Delta}+s}{2}}
\pFq{2}{1}{-s,\frac{d-p}{2}-1}{2-\frac{d-p}{2}-s}{\frac{x}{\bar{x}}}
\pFq{2}{1}{\widehat{\Delta},\frac{p}{2}}{\widehat{\Delta}-\frac{p}{2}+1}{x\bar{x}}\,.
\end{equation}
This concludes our discussion of defect two-point functions and their conformal block
decompositions, and we are now ready to take a first look at inversion formulas.

\subsection{Lightcone bootstrap and inversion formulas}
\label{sec:lcBootstrap}

As in the analysis of bulk four-point functions, the two different ways of evaluating the
defect two-point function, either through the bulk or the defect channel expansions, must
give the same answer, \ie
\begin{equation}\label{eq:2ptCrossing}
	\mathcal{F}(x, \bar{x}) =\left(\frac{(1-x)(1-\bar{x})}{(x \bar{x})^{\frac{1}{2}}}
\right)^{-\frac{\Delta_1+\Delta_2}{2}}
	\sum_{ \mathcal{O}} c_{12 \mathcal{O}}a_{\mathcal{O}}
	f_{\Delta,\ell}(x, \bar{x})
	= \sum_{\widehat{\mathcal{O}}} b_{1  \widehat{\mathcal{O}}}
b_{2 \widehat{\mathcal{O}}} \hat{f}_{\widehat{\Delta},s}(x, \bar{x})
	\,.
\end{equation}
This equation is the defect version of the famous crossing symmetry constraint for bulk
CFTs, and one would suspect that standard bootstrap techniques should
be applicable. Closer inspection shows, though, that equation \eqref{eq:2ptCrossing} does
not exhibit positivity: the bulk data is a product of two different constants. Although
innocent looking, this implies that the standard numerical techniques of \cite{Rattazzi:2008pe}
cannot be applied in this case (see \cite{Gliozzi:2013ysa,El-Showk:2016mxr} for alternative
proposals). It is possible however to study this equation analytically using lightcone
bootstrap techniques. The idea behind the original papers is the following: by taking a
special limit of the crossing equation it is possible to suppress operators on one of the
channels and therefore simplifying the equations significantly. The first relevant limit
is $(1-\bar{x}) \rightarrow 0$, where most of the bulk operators are suppressed and
only the bulk identity remains. The second limit is $x \rightarrow 0$, in this case the defect channel
is dominated by operators with small transverse twist.

The $(1-\bar{x}) \rightarrow 0$ limit was studied at length in \cite{Lemos:2017vnx} where
a Lorentzian inversion formula was obtained that captures the defect data as an analytic
function of the transverse spin. With the recent progress in our understanding of bulk
channel blocks and their relation to Calogero-Sutherland models, it is now possible to
complete the previous analysis and to present a Lorentzian inversion formula for the
bulk channel of defect two-point functions. This allows us to also study the second
limit $x \rightarrow 0$, see sections \ref{sec:lorentzianInversion} - \ref{sec:3dIsing_twist_defect}.
Before we get there, however, let us briefly review the known inversion formulas for
defect two-point functions.

\subsubsection{Euclidean inversion formula for the bulk channel}
\label{sec:EuclInversion}

Conformal blocks do not form a complete orthonormal set of functions but there exist
some closely related objects that do, the so-called Euclidean conformal partial waves. They are obtained as linear combination of a blocks and its shadow. For the bulk
channel blocks we introduced in equation \eqref{eq:fblock}, the Euclidean partial waves
read \cite{Isachenkov:2018pef}
\begin{equation} \label{eq:CPWbulk}
	F^E_{\Delta,\ell}(x,\bar x) = \frac12 \left(\, f_{\Delta,\ell}(x,\bar x) +
	\frac{K_{d-\Delta,\ell}}{K_{\Delta,\ell}}\,
	f_{d-\Delta,\ell}(x,\bar x)\right) \,.
\end{equation}
In order to obtain a complete orthonormal basis, we let $\Delta$ run through $\Delta
= d/2 + i \mathbb{R}^+$ and the coefficients are given by
\begin{equation}\label{eq:kappa}
	K_{\Delta,\ell} =  \frac{\Gamma\left(\Delta-p-1\right)}{\Gamma\left(\Delta-\frac{d}{2}\right)} \frac{ \Gamma\left( \frac{\Delta}{2} - \frac{1}{2} \right) }{\Gamma\left( \frac{\Delta}{2} - \frac{p}{2} - \frac{1}{2} \right)} \kappa_{\Delta+\ell} \,,\quad \kappa_\beta = \frac{ \Gamma\left( \frac{\beta}{2} \pm a \right)\Gamma\left( \frac{\beta}{2} \right) }{ 2\pi^2 \Gamma(\beta)\Gamma(\beta-1) } \,.
\end{equation}
By definition, the bulk channel conformal partial wave amplitude $c(\Delta,\ell)$ of
the defect two-point functions $\mathcal{F}$ is obtained through the following integral
transform 
\begin{align} \label{eq:Euclideaninversion}
c(\Delta,\ell) = \mathcal{N}_{\Delta,\ell} \int_{\mathbb{C}} d^2x \,
\mu(x,\bar x) \, F^E_{\Delta,\ell}(x,\bar x) \, \mathcal{F}(x,\bar x)\, .
\end{align}
where the integration region is the whole complex plane and the measure factor reads
\begin{align}\label{eq:inversionMeasure}
	\mu(x,\bar x) &=  \left(\frac{(1-x)(1-\bar{x})}{\sqrt{x\bar{x}}}\right)^{\frac{\Delta_1+\Delta_2}{2}} \frac{|x-\bar{x}|^{d-p-2}|1-x\bar{x}|^{p}}{\left[(1-x)(1-\bar{x})\right]^d}\,.
\end{align}
The normalization factor is given by,
\begin{equation}
	\mathcal{N}_{\Delta,\ell} = \frac{1}{2\pi} \frac{\Gamma\left( \ell + \frac{d-2}{2} \right)\Gamma\left( \ell + \frac{d}{2} \right)}{\Gamma\left( \ell + 1 \right)\Gamma\left( \ell + d-2 \right)} \frac{ \Gamma\left( \frac{\ell}{2} + \frac{1}{2} \right)\Gamma\left( \frac{\ell}{2} + \frac{d}{2} - \frac{1}{2} \right) }{ \Gamma\left( \frac{\ell}{2} + \frac{p}{2} + \frac{1}{2} \right)\Gamma\left( \frac{\ell}{2} + \frac{d-p}{2} - \frac{1}{2} \right) } \frac{K_{\Delta,\ell}}{K_{d-\Delta,\ell}} \,.
\end{equation}
Conversely, one may represent the two-point function $\mathcal{F}$ through the partial wave
amplitudes as a sum over even integer $\ell$ and an integral over conformal weights in the
range $\Delta = d/2 + i \mathbb{R}^+$. Splitting the partial waves into block and shadow,
we can re-express this decomposition of the two-point function in terms of Mellin-Barnes
integrations along the entire line $\Delta = d/2 + i \mathbb{R}$. If we assume that the
partial wave amplitudes fall off sufficiently fast and that they have poles only along
the real axis we can recover the following standard conformal block decomposition of the
defect two-point function
\begin{equation}
\mathcal{F}(x, \bar{x}) =\left(\frac{(1-x)(1-\bar{x})}{(x \bar{x})^{\frac{1}{2}}}
\right)^{-\frac{\Delta_1+\Delta_2}{2}}
\sum_{ \mathcal{O}} c_{12 \mathcal{O}}a_{\mathcal{O}}
f_{\Delta,\ell}(x, \bar{x}) \,,
\end{equation}
provided that the residues of the partial wave amplitudes are related to the coefficients
of the conformal block decomposition \eqref{eq:2ptbulkCBD} by
\begin{equation}\label{eq:resCDef}
c_{12 \mathcal{O}}a_{\mathcal{O}} = - \mathop{\mathrm{Res}}\limits_{\Delta'=\Delta}
c(\Delta',\ell) \qquad \text{($\Delta$ generic)} \, .
\end{equation}
Here we used the standard CFT conventions, see equation (5.19) of \cite{Isachenkov:2018pef}.
Equation \eqref{eq:Euclideaninversion} is often referred to as an Euclidean inversion formula,
but it is simply the definition of a partial wave transform that assigns a partial wave
amplitude to a correlation function. Remarkably, it is possible to obtain the same quantity
through an integral over a Lorentzian domain on which $x$ and $\bar x$ are real. We will
derive this Lorenztian inversion formula for the bulk channel of defect two-point functions
in section \ref{sec:lorentzianInversion}.

\subsubsection{Lorentzian inversion formula for the defect channel}
\label{sec:LorInversionDefectCh}

A very similar discussion applies to the defect channel of the defect two-point functions.
In this case the relevant blocks were given in equation \eqref{eq:defBlock} and one can also use
these to build a complete orthogonal set of defect conformal partial waves. As in the case of
the bulk channel, these defect partial waves can be used to define a defect partial wave
amplitude $b(\widehat{\Delta},s)$ that depends on the two parameters $\widehat{\Delta}$
and $s$ of the defect partial waves. The integration domain in the partial wave transform
is the Euclidean domain in which $x$ and $\bar x$ are complex conjugates, see
\cite{Lemos:2017vnx} for details. Defect partial wave amplitudes are expected to be
meromorphic with poles along the real line and residues
\begin{equation}
b_{\phi  \widehat{\mathcal{O}}}^2 =
- \mathop{\mathrm{Res}}\limits_{\widehat{\Delta}'=\widehat{\Delta}} b(\widehat{\Delta}',s)
\end{equation}
which reproduce the coefficients of the defect channel block decomposition \eqref{eq:defectCBD}
of the defect two-point function.

In this case the authors of \cite{Lemos:2017vnx} were able to reconstruct the defect channel
partial wave amplitude $b(\widehat{\Delta},s)$ through an integration over a Lorentzian domain.
More precisely, they proved the following Lorentzian inversion formula for the defect
channel,
\begin{equation}\label{eq:lorDefInversion}\begin{aligned}
\left. b(\widehat{\Delta},s) \right|_\text{poles} &= \int_0^1 \! \frac{dx}{2x} \,
x^{-\frac{\widehat{\tau}}{2}} \int_1^{\frac{1}{\bar{x}}} \! \frac{d\bar{x}}{2\pi i} \, (1-x\bar{x})(\bar{x}-x)\bar{x}^{-\frac{\widehat{\Delta}+s}{2}-2}
\pFq{2}{1}{s+1,2-\frac{d-p}{2}}{\frac{d-p}{2}+s}{\frac{x}{\bar{x}}} \\[2mm]
&\qquad \qquad \qquad \qquad \times
\pFq{2}{1}{1-\widehat{\Delta},1-\frac{p}{2}}{1+\frac{p}{2}-\widehat{\Delta}}{x\bar{x}}
\op{Disc}\mathcal{F}(x,\bar{x}) \,.
\end{aligned}\end{equation}
The integrand does not contain the defect two-point functions $\mathcal{F}$ itself, but
a certain \emph{discontinuity} thereof which is defined as
\begin{equation}
	\op{Disc}\mathcal{F}(x,\bar{x}) = \mathcal{F}^{\circlearrowleft}(x,\bar{x}) -
\mathcal{F}^{\circlearrowright}(x,\bar{x})  \,.
\end{equation}
Here, $\mathcal{F}^{\circlearrowleft}$ or $\mathcal{F}^{\circlearrowright}$ indicates
that $\bar{x}$ should be taken around $1$ in the direction shown, leaving $x$ fixed.

The symbol $|_\text{poles}$ in equation \eqref{eq:lorDefInversion} means that the partial wave amplitude
$b(\widehat{\Delta},s)$ and the integral on the right hand side possess poles in the
same positions with the same residues, but they are different functions otherwise. An
exact Lorentzian inversion formula for the defect channel partial wave amplitude is
also known, see \cite{Lemos:2017vnx}. The version we have displayed here is a result of some
simplifications in the exact integral expression that do, however, not effect the
poles and hence do not alter the dynamical content.

%% file: inversion.tex
%!TEX root = ../bulk_inversion.tex
%%%%%%%%%%%%%%%%%%%%%%%%%%%%%%%%%%%%%%%%%%%%%

\section{Lorentzian inversion formula for the bulk channel}
\label{sec:lorentzianInversion}

We now come to the main result of this paper, namely to our new  Lorentzian inversion formula for
the bulk channel of defect two-point functions. The formula is spelled out in the first subsection.
We support this formula through a detailed comparison with the Lorentzian inversion formula of
\cite{Caron-Huot:2017vep}. As we recalled above, the bulk channel blocks and partial waves for
defect two-point functions resemble those of bulk four-point functions, and there are some cases
in which they are essentially the same. This is most easily understood within the Calogero-Sutherland
approach to conformal blocks \cite{Isachenkov:2018pef} that was initiated in \cite{Isachenkov:2016gim}.
We will review the approach and give the complete list of relations between the two types of blocks
in the second subsection. Equipped with this knowledge we will then go through all the cases that
can be mapped to bulk four-point blocks and verify that our inversion formula coincides with the
one in \cite{Caron-Huot:2017vep}. Let us stress however, that there are many cases in which such
a map to Caron-Huot's formula is not possible. In this sense, our inversion formula is a true
extension.

\subsection{The main result}
\label{sec:mainResults}

As we have reviewed in the previous section, a Lorentzian inversion formula for the \textit{defect
channel} of a defect two-point function was derived in \cite{Lemos:2017vnx}. This inversion formula
allowed to extract information on defect operators from knowledge of  the bulk. Through a Lorentzian
inversion formula for the \textit{bulk channel} of the kind we shall state now, it is possible to go
in the other direction, \ie to infer properties of the bulk from information of the defect fields.

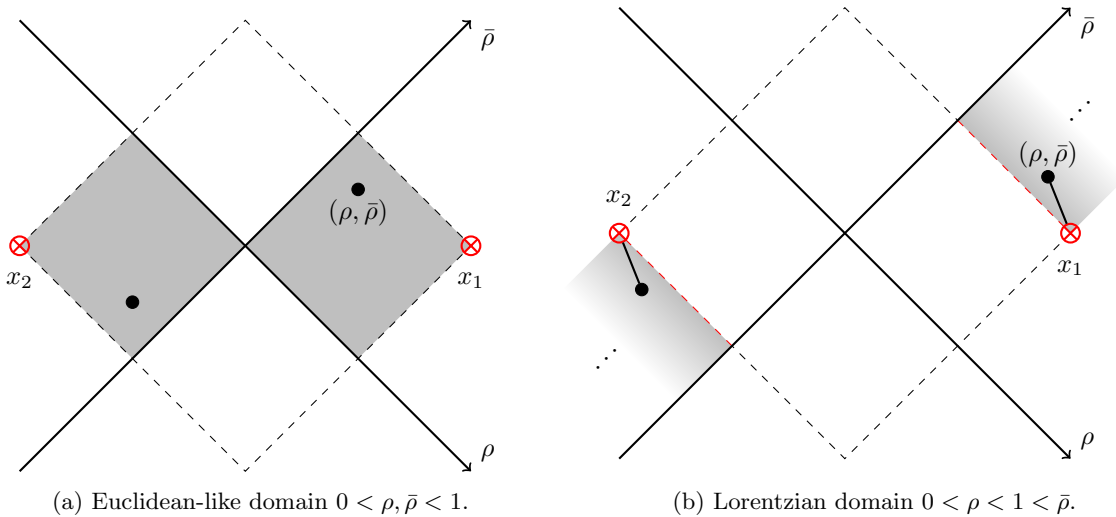
\begin{figure}
	\centering
	\mbox{\begin{subfigure}[t]{0.45\textwidth}
			\mbox{\begin{tikzpicture}[scale = 3, cross/.style={path picture={
						\draw[red]
						(path picture bounding box.south east) -- (path picture bounding box.north west) (path picture bounding box.south west) -- (path picture bounding box.north east);
				}}]
				\fill[lightgray] (-1,0) -- (-.5,.5) -- (0,0) -- (-.5,-.5) -- cycle;
				\fill[lightgray] (1,0) -- (.5,.5) -- (0,0) -- (.5,-.5) -- cycle;
				
				\draw[dashed] (-1,0) -- (0,1) -- (1,0) -- (0,-1) -- cycle;
				
				\draw[thick, ->] (-1,-1) -- (1,1) node[below right] {$\bar{\rho}$};		
				\draw[thick, ->] (-1,1) -- (1,-1) node[above right] {$\rho$};

				\node[draw, red, circle, cross, scale = .75, thick, label={[label distance=.1cm]below:$x_2$}] (x2) at (-1,0) {};
				\node[draw, red, circle, cross, scale = .75, thick, label={[label distance=.1cm]below:$x_1$}] (x1) at (1,0) {};
				
				\node[draw, circle, fill, scale = .5] (x4) at (.5,.25) {};
				\node[below] at (x4) {$(\rho,\bar{\rho})$};
				\node[draw, circle, fill, scale = .5] (x3) at (-.5,-.25) {};
				\end{tikzpicture}}
			\caption{Euclidean-like domain  $0 < \rho,\bar{\rho} < 1$.}
			\label{fig:rhoRegionsEucl}
	\end{subfigure}}
	\hfill
	\mbox{\begin{subfigure}[t]{0.5\textwidth}
			\mbox{\begin{tikzpicture}[scale = 3, cross/.style={path picture={
						\draw[red]
						(path picture bounding box.south east) -- (path picture bounding box.north west) (path picture bounding box.south west) -- (path picture bounding box.north east);
				}}]
				\shade[rotate = 45, top color = lightgray, bottom color = white, transform canvas = {rotate = -45}] (-1,0) -- (-1.25,-.25) -- (-.75,-.75) -- (-.5,-.5) -- cycle;
				\node[rotate = 45] at (-1.05,-.55) {\dots};
				\shade[rotate = 45, top color = white, bottom color = lightgray, transform canvas = {rotate = -45}] (1,0) -- (1.25,.25) -- (.75,.75) -- (.5,.5) -- cycle;
				\node[rotate = 45] at (1.05,.55) {\dots};
				
				\draw[dashed] (-1,0) -- (0,1) -- (.5,.5) (1,0) -- (0,-1) -- (-.5,-.5);
				\draw[dashed, red] (.5,.5) -- (1,0) (-.5,-.5) -- (-1,0);
				
				\draw[thick, ->] (-1,-1) -- (1,1) node[below right] {$\bar{\rho}$};		
				\draw[thick, ->] (-1,1) -- (1,-1) node[above right] {$\rho$};

				\node[draw, red, circle, cross, scale = .75, thick, label={[label distance=.1cm]above:$x_2$}] (x2) at (-1,0) {};		
				\node[draw, red, circle, cross, scale = .75, thick, label={[label distance=.1cm]below:$x_1$}] (x1) at (1,0) {};
				
				\node[draw, circle, fill, scale = .5] (x4) at (.9,.25) {};
				\node[above] at (x4) {$(\rho,\bar{\rho})$};
				\node[draw, circle, fill, scale = .5] (x3) at (-.9,-.25) {};
				\draw[thick] (x3) -- (x2);
				\draw[thick] (x4) -- (x1);
				\end{tikzpicture}}
			\caption{Lorentzian domain $0 < \rho < 1 < \bar{\rho}$.}
			\label{fig:rhoRegionsLor}
	\end{subfigure}}
	\caption{Domains in Lorentzian kinematics. The spherical defect intersects the $(\rho,\bar{\rho})$-plane at $x_1$ and $x_2$. In order to pass from the Euclidean-like region \subref{fig:rhoRegionsEucl} (depicted in gray) to the Lorentzian region \subref{fig:rhoRegionsLor}, we need to cross two light-cones, once of $x_1$ and once of $x_2$. Time-like distances are depicted by a bold line.}
	\label{fig:rhoRegions}
\end{figure}

The geometrical setup for the Lorentzian domain is depicted in figure \ref{fig:rhoRegions}. This setup is most conveniently described by the coordinates \cite{Lauria:2017wav}
\begin{equation}
	\rho = \frac{1-\sqrt{x}}{1+\sqrt{x}} \,,\quad \bar{\rho} = \frac{1-\sqrt{\bar{x}}}{1+\sqrt{\bar{x}}} \,,
\end{equation}
where the defect is now
spherical, intersecting the $(\rho,\bar{\rho})$-plane in two points. The configuration is analogous to the
one that describes a four-point function of local operators \cite{Hogervorst:2013sma}, and was used in \cite{Caron-Huot:2017vep} to derive the Lorentzian inversion formula.

With all the preparation in the previous section, and in particular through our discussion of the
Euclidean inversion formula in subsection \ref{sec:EuclInversion}, it is not difficult to state the
Lorentzian inversion formula for the bulk channel of defect two-point functions. In fact, we claim
that the bulk channel partial wave amplitude $c(\Delta, \ell)$ takes the form
\begin{equation}
	c(\Delta,\ell) = c^t(\Delta,\ell) + (-1)^\ell c^u(\Delta,\ell)\,,
\end{equation}
where the $t$-channel term is given by
\begin{equation}\label{eq:defInversion}\boxed{ \qquad
	c^t(\Delta,\ell) = \frac{\kappa_{\Delta+\ell}}{4} \int_0^1 d^2x \, \mu(x,\bar{x})
f_{\ell+d-1,\Delta-d+1}(x,\bar{x}) \operatorname{dDisc} \mathcal{F}(x,\bar{x}) 
	\qquad }
\end{equation}
and the $u$-channel term is of the same form, but with the two local bulk fields exchanged.
In this formula, the factors $\kappa$ and $\mu$ are the same as 
in the Euclidean setup (equations \eqref{eq:kappa} and \eqref{eq:inversionMeasure}, respectively) and $f$ is a bulk channel block of the defect two-point
function that we constructed in equation \eqref{eq:fblock}. The only new ingredient here is the
\emph{double discontinuity} which we define as
\begin{equation}
\operatorname{dDisc} \mathcal{F}(x,\bar{x}) = \cos(\pi a) \mathcal{F}(x,\bar{x}) -
\frac12 e^{-i\pi\frac{\Delta_1+\Delta_2}{2}} \mathcal{F}^{\circlearrowleft}(x,\bar{x})
- \frac12 e^{+i\pi\frac{\Delta_1+\Delta_2}{2}} \mathcal{F}^{\circlearrowright}(x,\bar{x}) \,,
\end{equation}
where $\mathcal{F}^{\circlearrowleft}$ or $\mathcal{F}^{\circlearrowright}$ now means that
$\bar{x}$ should be taken around $0$ in the direction shown, leaving $x$ fixed. Note that
this differs from the prescription in the defect channel inversion formula.

Even though the
double discontinuity has a definite sign for identical scalars (\ie $a=0$), we do not expect
this property to hold for the coefficients $c_{12 \mathcal{O}}a_{\mathcal{O}}$ of the bulk channel
block decomposition in general. The sign is rather determined by the function $\mathcal{F}$.

\subsection{Calogero-Sutherland approach to the four-point and bulk OPE blocks}
\label{sec:CSapproach}

In order to establish our new inversion formula we shall need some background on the relation
between bulk channel blocks for defect two-point functions and the blocks of scalar four-point
functions. Some special cases of this relation have already appeared in the literature \cite{Billo:2016cpy,Gadde:2016fbj,Liendo:2016ymz}, but the analysis
was only completed in \cite{Isachenkov:2018pef}, based on a detailed comparison of the relevant
Casimir equations, including their boundary conditions. What made it systematic was the observation
that Casimir equations for conformal blocks can always be mapped to eigenvalue equations of some
Calogero-Sutherland Hamiltonian \cite{Isachenkov:2016gim}. The latter provides a standard
representation of Casimir equations that makes relations manifest.

All Casimir equations that are relevant for the kind of blocks we study in this paper can be
mapped to the eigenvalue equations for the following Calogero-Sutherland Hamiltonian,
\begin{align}\label{eq:HCS}
H_\textrm{CS} &= -\sum_{i=1}^{2} \frac{\partial^2}{\partial \tau_i^2} +
\frac{k_3(k_3-1)}{2}\left[ \sinh^{-2}\left(\frac{\tau_1+\tau_2}{2}\right)+
\sinh^{-2}\left(\frac{\tau_1-\tau_2}{2}\right) \right]
\notag\\[2mm]
&\qquad
+ \sum_{i=1}^{2}\left[ k_2(k_2-1)\sinh^{-2}\left(\tau_i\right) +
\frac{k_1(k_1+2k_2-1)}{4}\sinh^{-2}\left(\frac{\tau_i}{2}\right) \right] \ .
\end{align}
In applications to the bulk channel of defect two-point blocks, the coordinates $\tau_i$ of
the Calogero-Sutherland Hamiltonian are related to the cross ratios $x$, $\bar{x}$ we defined
in equation \eqref{eq:xxbar} through
$$ x = \tanh^{-2} \frac{\tau_1 + \tau_2}{4} \quad , \quad
\bar x = \tanh^{-2} \frac{\tau_1 - \tau_2}{4} \ . $$
To construct Euclidean partial waves and blocks, the Hamilton operator \eqref{eq:HCS} is to be
considered on a subspace of the semi-infinite hypercuboid that is parametrized by the coordinates
\begin{equation}\label{eq:tauRange}\
\tau_1 \in [0,\infty[ \,,\quad \tau_2 \in i[0,2\pi] \,.
\end{equation}
The parameters $k_a$ depend on the precise setup, \ie on the defect dimension $p$, the weights
of the external scalars, and the dimension $d$ through,
\begin{align} \label{eq:Casp0par}
\quad k_1=\frac{d}{2}-p-1 \,,\quad k_2=\frac{p}{2} \,,\quad k_3=\frac12 + a \,.
\end{align}
The wave functions $\psi(\tau)$ are defined through the following Schr\"odinger equation
\begin{equation} \label{eq:schrodinger}
H_{CS}\psi_\epsilon(\tau) = \epsilon \psi_\epsilon(\tau)\,,\quad \epsilon =
-\frac14 C_{\Delta,\ell} - \frac{d^2-2d+2}{8} \,,
\end{equation}
with eigenvalues of the form
\begin{equation}
C_{\Delta,\ell} = \Delta(\Delta-d)+\ell(\ell+d-2) \,,
\end{equation}
where $\Delta$ and $\ell$ denote the conformal weight and the spin of the intermediate
field, respectively. There are eight independent solutions corresponding to the symmetries
of the Hamiltonian \eqref{eq:HCS}. These solutions are related by
\begin{equation}
\Delta \leftrightarrow d-\Delta \,,\quad \ell \leftrightarrow 2-d-\ell \,,\quad \Delta
\leftrightarrow 1-\ell \,,
\end{equation}
and they are distinguished by their asymptotics \eqref{eq:defblocknormxxb}. These eight
solutions are related to the Harish-Chandra (or pure) functions we defined in equation
\eqref{eq:HSfct} as
\begin{equation}\label{eq:cb_wavefct}
\cblock{f^{HS}}{p,a,d}{\Delta,\ell}{x,\bar{x}} =  2^{2\Delta-d+1} \omega(\tau)\psi_\epsilon(\tau) \,.
\end{equation}
The ``gauge transformation''
$\omega(\tau)$  is given by
\begin{align}\label{eq:gauge1}
\omega (\tau) &= \prod_{i=1}^{2}\sinh^{-\frac{d}{2}+\frac{p}{2}+1}\left(\frac{\tau_i}{2}\right)
\cosh^{-\frac{p}{2}}\left(\frac{\tau_i}{2}\right) \sinh^{-\frac12}\left(\frac{\tau_1 \pm
\tau_2}{2}\right).
\end{align}
In the final factor we used the standard convention $h(x\pm y) = h(x+y) h(x-y)$. In order to
construct the true (regular) wave function of the Calogero-Sutherland model on the domain
\eqref{eq:tauRange}, one has to take a special linear combination of the eight Harish-Chandra
functions. The physical wave functions of the Calogero-Sutherland model are related to the
Euclidean partial waves $F^E$. Conformal blocks \eqref{eq:fblock} arise from a decomposition
of the partial waves into block and its shadow as described in equation \eqref{eq:CPWbulk}.
\medskip

The theory of bulk four-point blocks possesses a similar description, see \cite{Isachenkov:2016gim,
Isachenkov:2017qgn}. In this case the relevant Casimir equations are known to take the form
\begin{equation}
L^2 = \frac12 H'_{CS} + \epsilon_0 \,,\quad \epsilon_0 = \frac{d^2-2d+2}{8}
\end{equation}
with parameters $k_a$
\begin{align}\label{eq:Cas00par}
 k_1= -2b \,,\quad k_2=a+b+\frac12 \,,\quad k_3=\frac{d-2}{2} \, ,
\end{align}
that are determined by the conformal weights of the external scalar fields through the
standard combinations $2a= \Delta_2-\Delta_1$ and $2b=\Delta_3-\Delta_4$. We placed a
prime $'$ on the Hamiltonian to indicate that it actually depends on two variables $u_1$
and $u_2$ that are complex conjugates of each other, and belong to the range
\begin{equation} \label{eq:urange}
\Re u_i \in [0, \infty[ \,, \quad \Im u_1 = - \Im u_2 \in [0,\pi[\ .
\end{equation}
As in the previous case, (Harish-Chandra) eigenfunctions of this Calogero-Sutherland
Hamiltonian differ from the bulk four-point blocks $g_{\Delta,\ell}$ by a gauge
transformation. The latter now takes the form
\begin{align}\label{eq:gauge2}
\omega'(u_1,u_2) &= \prod_{i=1}^{2} \sinh^{a+b-\frac12}\left(\frac{u_i}{2}\right)
\cosh^{-(a+b)-\frac12}\left(\frac{u_i}{2}\right) \sinh^{-\frac{d-2}{2}}
\left(\frac{u_1\pm u_2}{2}\right) \,,
\end{align}
and the eigenvalues $\epsilon'$ of the Calogero-Sutherland Hamiltonian $H'$ are
related to the conformal weight $\Delta$ and the spin $\ell$ of the intermediate
field by $\epsilon'=-\frac12 C_{\Delta,\ell} - 2\epsilon_0$. We will denote the
bulk four-point block by,
\begin{equation}
g_{\Delta,\ell}(z,\bar{z}) \equiv \cblock{g}{a,b,d}{\Delta,\ell}{z,\bar{z}}\,,
\end{equation}
where
\begin{equation}
z = -\sinh^{-2}\left( \frac{u_1}{2} \right) \,,\quad \bar{z} = -\sinh^{-2}
\left( \frac{u_2}{2} \right)\,.
\end{equation}
This concludes our brief discussion of the Calogero-Sutherland approach to both
bulk chanel blocks for defect two-point functions and to four-point blocks.
\medskip

In order to obtain relations between the two types of blocks we note that the parameters
$k_a$ of the Calogero-Sutherland potential are not uniquely defined, \ie different choices
can give rise to identical Casimir equations. This is partly due to the fact that the
parameters appear quadratically in the potential. In addition, one may show that a
simultaneous shift of the coordinates $\tau_i  \rightarrow \tau_i + i\pi$ for $i=1,2$
leads to a Calogero-Sutherland Hamiltonian  of the form \eqref{eq:HCS} with different
parameters. The complete list of symmetries is given in table \ref{tab:sym}. These
innocent looking replacements have remarkable  consequences, since they produce
non-trivial relations between the blocks of various (defect) configurations. In
particular, they relate certain bulk channel blocks to four-point blocks. However,
in doing so one has to make sure that the range \eqref{eq:tauRange} is actually
mapped to the range \eqref{eq:urange}.\footnote{See \cite{Isachenkov:2018pef} for a
very detailed explanation. Let us note that the symmetry $\tilde{\sigma}$ changes
the range and hence the analytic structure of the conformal blocks.}
\begin{table}[h!]
	\centering
	\caption{Symmetries of the Calogero-Sutherland model for generic values of the
		parameters. The last symmetry also involves a shift $ \tau'_i = \tau_i \pm
		i \pi$ of the coordinates.}
	\begin{tabular}{l|c|c|c}
		& $p'$ & $a'$ & $d'$ \\[2mm] \hline & & & \\[-3mm]
		$\varrho_1$ & $p$ & $a$ & $2p-d+6$ \label{eq:k1_blkdef} \\[2mm]
		$ \varrho_2$ & $2-p$ & $a$  & $8-d$ \label{eq:k2_blkdef} \\[2mm]
		$\varrho_3$ & $ p $ & $-a$ & $ d $ \label{eq:k3_blkdef} \\[2mm]
		$ \tilde \varrho$ & $ 4-d+p $ & $ a $ & $ 8-d $ \label{eq:marco_blkdef}
	\end{tabular}
	\label{tab:sym}
\end{table}
Along these lines we obtain two dualities involving the symmetry $\sigma_1$ from
table \ref{tab:sym},
\begin{align} \label{eq:genrel3}
\cblock{f}{p=0,a,d}{\Delta,\ell}{x,\bar{x}} &= (x\bar{x})^{\frac{a}{2}} \cblock{g}{a,0,d}{\Delta,\ell}{1-x,1-\bar{x}} \,, \\[2mm]
\cblock{f}{p=2,a,d}{\Delta,\ell}{x,\bar{x}} &= \frac{(1-x)(1-\bar{x})}{1-x\bar{x}} (x\bar{x})^{\frac{a}{2}}
\cblock{g}{a,0,d-2}{\Delta-1,\ell+1}{1-x,1-\bar{x}} \,,
\label{eq:genrel4}
\end{align}
and two dualities involving the symmetry $\sigma_2$,
\begin{align}  \label{eq:genrel1}
\cblock{f}{\frac{d}{2}-1,a,d}{\Delta,\ell}{x,\bar{x}} &= (-1)^{-\frac{\ell}{2}} 2^{\Delta} (y\bar{y})^{-\frac14}
\cblock{g}{-\frac14+\frac{a}{2},-\frac14-\frac{a}{2},\frac{d}{2}+1}{\frac{\Delta+1}{2}, \frac{\ell}{2}}{y,\bar{y}} \\[2mm]
&= 2^{\Delta} (\gamma\bar{\gamma})^{-\frac14} \left[(1-\gamma)(1-\bar\gamma)\right]^{-\frac{a}{2}}
\cblock{g}{\frac14-\frac{a}{2},-\frac14-\frac{a}{2},\frac{d}{2}+1}{\frac{\Delta+1}{2}, \frac{\ell}{2}}{\gamma,\bar{\gamma}}
\,, \\[2mm]
\cblock{f}{p,a,d=4}{\Delta,\ell}{x,\bar{x}} &= (-1)^{-\frac{\ell-p+1}{2}} 2^{\Delta} (y\bar{y})^{-\frac14}
\left| \sqrt\frac{y-1}{y} - \sqrt\frac{\bar{y}-1}{\bar{y}} \right|^{p-1} \notag
\\[2mm]  &\qquad\times
\cblock{g}{-\frac14+\frac{a}{2},-\frac14-\frac{a}{2},p+2}{\frac{\Delta+p}{2}, \frac{\ell-p+1}{2}}{y,\bar{y}}
\, ,
\label{eq:genrel2}
\end{align}
where
\begin{equation}
\gamma = \frac{y}{y-1} = \left(\frac{1-x}{1+x}\right)^2 \,,\quad
\bar{\gamma} = \frac{\bar{y}}{\bar{y}-1} = \left(\frac{1-\bar{x}}{1+\bar{x}}\right)^2 \,.
\end{equation}
We will use these dualities to verify our Lorentzian inversion formula for the bulk channel. Let us stress that the list of relations between bulk channel blocks $f$ and four-point blocks $g$ is exhaustive, if we demand that Euclidean domains are mapped to Euclidean domains. More general relations are possible if one allows for analytic continuations, see the discussion around equation (2.23) in \cite{Isachenkov:2018pef}.

\subsection{Derivation of the Lorentzian inversion formula}
\label{sec:derivation}

It should be possible to derive our Lorentzian inversion formula \eqref{eq:defInversion} for the bulk 
channel following the same steps as in \cite{Caron-Huot:2017vep}, starting from 
the characterization of the partial wave amplitude $c(\Delta,\ell)$ in equation \eqref{eq:Euclideaninversion}.
However, the close relation between defect two-point functions and bulk four-point functions that we 
exposed through the relation with Calogero-Sutherland models and their solution theory, makes it 
possible to come up with an ansatz right away, namely 
\begin{equation}
		c^t(\Delta,\ell) \propto \int_0^1 \! d^2x \, \mu(x,\bar{x})  
\cblock{f}{p,a,d}{\ell+d-1,\Delta-d+1}{x,\bar{x}} \operatorname{dDisc} \mathcal{F}(x,\bar{x}) \,.
\end{equation}
Our strategy here is to verify this ansatz and fix the prefactor by using the dualities between bulk channel 
defect blocks and conformal blocks for scalar four-point functions described in subsection \ref{sec:CSapproach}.
Let us first address the case of a defect with $p=0$. As discussed in \cite{Gadde:2016fbj,Isachenkov:2018pef}, such 
a pointlike defect is localized in two points $x_3 = y_1$ and $x_4 = y_2$ and, up to the different choice of 
normalizations, the defect two-point function for $p=0$ looks like a four-point function of the form 
$$ \langle \phi_0(y_1) \phi_0(y_2) \phi_1(x_1) \phi_2(x_2) \rangle = \frac{1}{|y_1-y_2|^{2\Delta_0} 
|x_1-x_2|^{\Delta_1 + \Delta_2}} [(1-z)(1-\bar{z})]^\frac{a}{2} \mathcal{G}(z,\bar{z})\ ,  $$ 
where we assigned some conformal weight $\Delta_0 = \Delta_3 = \Delta_4$ to the two fictitious new 
fields. We also  used standard conventions to prepare a function $\mathcal{G}$ of the two cross-ratios 
$$  u = \frac{(x_1-x_2)^2(y_1-y_2)^2}{(x_1-y_1)^2(x_2-y_2)^2} = z \bar{z} \ ,\quad 
    v = \frac{(x_1-y_2)^2(x_2-y_1)^2}{(x_1-y_1)^2(x_2-y_2)^2} = (1-z) (1-\bar{z})\, , $$
from the insertion points of the four fields. Taking into account that our defect operators are normalized
such that $\langle \mathcal{D}^{(0)} \rangle = 1$, we arrive at the following relation 
\begin{eqnarray*} 
 [(1-z)(1-\bar{z})]^\frac{a}{2} \mathcal{G}(z,\bar{z}) & = & 
 |x_1-x_2|^{\Delta_1 + \Delta_2} \,  |y_1-y_2|^{2\Delta_0} \langle \phi_0(y_1) 
 \phi_0(y_2) \phi_1(x_1) \phi_2(x_2) \rangle\\[2mm] 
& = & |x_1-x_2|^{\Delta_1 + \Delta_2} \, \langle \mathcal{D}^{(0)} 
\phi_1(x_1) \phi_2(x_2) \rangle = \left(\frac{(1-x)(1-\bar{x})}
{\sqrt{x\bar{x}}}\right)^{\frac{\Delta_1+\Delta_2}{2}} \mathcal{F}(x,\bar{x})\ . 
\end{eqnarray*} 
In the last step we inserted the definition \eqref{eq:2ptFct} of the function $\mathcal{F}$ 
for $p=0$. With this relation between the functions $\mathcal{F}$ and $\mathcal{G}$ we can now
depart from the Lorentzian inversion formula for bulk four-point functions \cite{Caron-Huot:2017vep}, 
replace $\mathcal{G}$ by $\mathcal{F}$ and the four-point block $g$ by the associated bulk channel 
defect block to obtain 
\begin{equation}\label{eq:defInversion0}\begin{aligned}
c^t(\Delta,\ell) &= \frac{\kappa_{\Delta+\ell}}{4} \int_0^1 d^2x \, \left.\mu(x,\bar{x})\right|_{p=0} \cblock{f}{p=0,a,d}{\ell+d-1,\Delta-d+1}{x,\bar{x}} \operatorname{dDisc} \mathcal{F}(x,\bar{x}) \,.
\end{aligned}\end{equation}
This agrees with our Lorentzian inversion formula \eqref{eq:defInversion} if we specialize to $p=0$. 

In order to verify our formula for larger defect dimensions $p > 0$, we note that in Caron-Huot's 
derivation it did not matter that the function $\mathcal{G}$ was an actual four-point amplitude, as 
long as there exists a decomposition into four-point conformal blocks. Using the dualities reviewed
in subsection \ref{sec:CSapproach}, see equations \eqref{eq:genrel3} to \eqref{eq:genrel2}, we can write down an 
effective four-point amplitude $\mathcal{G}'$ for the defect amplitude $\mathcal{F}$ whose coefficients 
may then be determined by the Lorentzian inversion formula of \cite{Caron-Huot:2017vep}. All our dualities 
have the schematic form  
\begin{equation}
\cblock{f}{p,a,d}{\Delta,\ell}{x, \bar{x}} = d'_{\Delta,\ell} \omega(z',\bar{z}')
\cblock{g}{a',b',d'}{\Delta',\ell'}{z', \bar{z}'} \,,
\end{equation}
where the prefactor factorizes into a coordinate dependent part $\omega(z',\bar{z}')$, and a constant part 
that may dependent on the quantum numbers of exchanged operator $\mathcal{O}_{\Delta,\ell}$. Note that the 
precise relation between the parameters and the cross-ratios on both sides differs from case to case. 
Plugging such a relation between blocks into the bulk channel conformal block decomposition \eqref{eq:2ptbulkCBD} 
we obtain 
\begin{equation}\begin{aligned}
&\langle \mathcal{D}^{(p)}  \phi_1(x_1) \phi_2(x_2) \rangle =
|x_1-x_2|^{-(\Delta_1+\Delta_2)} \sum_{k} c_{12k}a_{k} d'_k \omega(z',\bar{z}')
\cblock{g}{a',b',d'}{\Delta'_k,\ell'_k}{z', \bar{z}'} \\
&\qquad = |x_1-x_2|^{-(\Delta_1+\Delta_2)} \omega(z',\bar{z}') \sum_{k} c_{12k}c'_{34k}
\cblock{g}{a',b',d'}{\Delta'_k,\ell'_k}{z', \bar{z}'} \\
&\qquad = |x_1-x_2|^{-(\Delta_1+\Delta_2)} \omega(z',\bar{z}') \mathcal{G}'(z',\bar{z}') \,,
\end{aligned}\end{equation}
where $c'_{34k} \equiv a_{k} d'_k$. Using the Lorentzian inversion formula of \cite{Caron-Huot:2017vep}, we can 
construct a conformal partial wave amplitude $c$ whose residues are given by the products $c_{12k}c'_{34k}$ and 
hence by $c_{12k}a_{k}$. Once all the details are filled in, each of the cases we listed in equations 
\eqref{eq:genrel3} to \eqref{eq:genrel2} above yields a special case of our formula \eqref{eq:defInversion}.
Since the analytic structure of the conformal blocks does not depend on the defect dimension $p$, see 
\cite{Isachenkov:2018pef}, we are confident that equation \eqref{eq:defInversion} does generalizes to any $p$, 
beyond the cases $p=0$, $2$ and $\frac{d}{2}-1$ for which dualities between bulk and defect blocks exist.

In \cite{Caron-Huot:2017vep}, positivity of the OPE coefficients was used to bound the Regge limit of the four-point correlator, \ie the limits $w \rightarrow 0+i0$ and $w\rightarrow \infty$, where $w$ is defined by
\begin{equation}
	\rho = rw \,,\quad \bar{\rho} = rw^{-1} \,.
\end{equation}
Without positivity at hand, we cannot prove such a behavior for the defect correlator. We will therefore assume, similar to what was done in \cite{Lemos:2017vnx}, that two-point functions exhibit a power-law behavior,
\begin{equation}
	\left(\frac{(1-x)(1-\bar{x})}{(x \bar{x})^{\frac{1}{2}}}
	\right)^{\frac{\Delta_1+\Delta_2}{2}} \mathcal{F}(x,\bar{x}) \lesssim w^{1-\ell_*} \,, \quad\text{as $w\rightarrow0$} \,, \quad\text{(similar for $w\rightarrow\infty$)}\, ,
\end{equation}
for some value $\ell_*$, which implies that our formula is valid for $\ell>\ell_*$. The actual value of $\ell_*$ depends on the theory and the correlator one considers. It would be interesting to prove a theory-independent lower bound on $\ell_*$ from first principles.

%% file: defect_identity.tex
%!TEX root = ../bulk_inversion.tex
%%%%%%%%%%%%%%%%%%%%%%%%%%%%%%%%%%%%%%%%%%%%%

\section{Inverting the defect identity}
\label{sec:lcDef}

Through the evaluation in the lightcone limit, Lorentzian inversion formulas have been applied to
obtain results on the large spin behavior of conformal weights and OPE coefficients.
We shall consider two such applications of our new inversion formula \eqref{eq:defInversion}. In
this section we are going to make a (nearly) model independent prediction for the one-point functions $a_\mathcal{O}$ of certain bulk operators $\mathcal{O}$ in the large spin limit.\footnote{This analysis can also be performed using standard lightcone bootstrap techniques \cite{llms}.} The
operators $\mathcal{O}$ to which our formula applies are those of twist $\tau = 2 \Delta_\phi$
that appear in the OPE of two identical scalar bulk fields $\phi$ with
conformal weight $\Delta_\phi$. The only assumptions that go into our analysis is that the
defect is chosen such that the expansion \eqref{eq:defectCBD} contains one term from the defect
identity $\widehat{\mathcal{O}} = 1$, and that all other terms are separated from this one by a
positive finite twist gap $\delta > 0$, i.e.\ $ \ \widehat{\tau}_k = \widehat{\Delta}_k - s_k
\geq \delta$ for all defect operators $\widehat{\mathcal{O}}_k$ that contribute to the
defect channel conformal block decomposition.

Unlike the bulk channel identity, the defect identity is not always present in the expansion. 
If $\phi$ is the lowest scalar of the
$3d$ Ising model, for example, then the defect identity $\widehat{\mathcal{O}}=1$ of the twist
defect, or any other defect that preserves the $\mathbb{Z}_2$ symmetry of the bulk Ising model,
cannot appear in the defect channel block expansion, because the field $\phi$ is odd under the
$\mathbb{Z}_2$ symmetry while the identity field is certainly even. A second example that
illustrates the issue arises from the study of surface critical phenomena in the $3d$ Ising
model. As is well known, there exists three possible \textit{surface} critical phenomena
which have been dubbed the ordinary, special, and extraordinary transitions \cite{cardy}.
For the extraordinary transition the spin field diverges near the boundary, and hence this
condition breaks the $\mathbb{Z}_2$ symmetry while the other two do not. Consequently, the
bulk spin field can couple to the identity of the defect/surface only in the case if the
extraordinary boundary condition. For the bootstrap program applied to boundary CFT see \cite{Liendo:2012hy,Gliozzi:2015qsa,Rastelli:2017ecj,Bissi:2018mcq,Mazac:2018biw}.

 Also our second requirement of a positive non-vanishing defect twist gap $\delta$ could be
violated for some defects. Here we shall give a slightly degenerate example with a defect of
dimension $p=0$ that nevertheless highlights the potential issue quite well. Every bulk
field $\phi$ gives rise to a pointlike defect that amounts to inserting two fields
at the two locations of the defect. The corresponding defect operator $\mathcal{D}^{(0)}$
takes the form
\begin{equation}\label{eq:pointlikeDphi}
\mathcal{D}^{(0)}_\phi = |y_{12}|^{2\Delta_\phi} \phi_1(y_1) \phi(y_2)\, .
\end{equation}
The $y$-dependent prefactor ensures that the defect operator is correctly normalized, i.e.\
that $\langle \mathcal{D}^{(0)}_\phi \rangle = 1$. Now we want to probe this pointlike defect
through a defect two point function in which we insert the same field $\phi$ at two points
$x_1$ and $x_2$ in the bulk, i.e.\ we consider
\begin{equation}
\langle \mathcal{D}^{(0)}_\phi \phi(x_1) \phi(x_2) \rangle = |y_{12}|^{2\Delta_\phi}
\langle \phi(y_1) \phi(y_2) \phi(x_1) \phi(x_2) \rangle\, .
\end{equation}
This equation relates the defect channel conformal partial wave decomposition \eqref{eq:defectCBD}
for the pointlike defect created by $\mathcal{D}^{(0)}_\phi$, to the $t$-channel expansion of the
bulk four-point function on the right-hand side. In a unitary theory, the latter starts with a
contribution from the insertion of the bulk identity.
This term comes with the factor $|y_{12}|^{-2\Delta_\phi}$, but for the identification with the defect two-point function it gets multiplied with the factor $|y_{12}|^{\Delta_\phi}$.
Hence, we conclude that the leading term in the defect
channel partial wave decomposition scales with a weight $\widehat{\Delta} = -\Delta_\phi$. In other
words, the twist of the leading twist field on the defect is given by $\widehat{\tau}_\text{min} = - 
\Delta_\phi$, which is actually negative.   
\medskip

After these words of caution, let us now start to analyse our bulk channel Lorentzian inversion
formula \eqref{eq:defInversion} in the lightcone limit when $1-x,\bar{x} \ll 1$. Following a
similar analysis in \cite{Caron-Huot:2017vep}, we note that the pole contributions of the
bulk channel partial wave amplitude can be computed through
\begin{equation}\label{eq:genFctExp}
\left.c^t(\Delta,\ell)\right|_{poles} \simeq \int_0^1\! \frac{dx}{2(1-x)} (1-x)^\frac{\ell-\Delta}{2}
\sum_{m=0}^{\infty} (1-x)^m \sum_{k=-m}^{m} B^{(m,k)}_{\Delta,\ell} C^t(x,\Delta+\ell+2k) \,,
\end{equation}
where $C^t(x,\beta)$ is the \emph{generating function},
\begin{equation}\label{eq:genFctDef}
C^t(x,\beta) = \left(\frac{1-x}{\sqrt{x}}\right)^{\Delta_\phi} \kappa_\beta \int_{0}^{x}\! \frac{d\bar{x}}{(1-\bar{x})^2}\,  \left(\frac{1-\bar{x}}{\sqrt{\bar{x}}}\right)^{\Delta_\phi} k_\beta(1-\bar{x}) \operatorname{dDisc} \mathcal{F}(x,\bar{x}) \,.
\end{equation}
Remarkably, the generating function is independent of the defect dimension $p$. The dependence on $p$ enters
the partial wave amplitude $c(\Delta,\ell)$ only through the coefficients $B^{(m,k)}_{\Delta,\ell}$. These
coefficients may be determined by expanding the kernel in equation \eqref{eq:defInversion},
\begin{equation}\begin{aligned}
	&(1-x)(1-\bar{x})^2\mu(x,\bar{x})f^\textit{HS}_{\ell+d-1,\Delta-d+1}(x,\bar{x}) \\
	&\qquad \equiv \left(\frac{(1-x)(1-\bar{x})}{\sqrt{x\bar{x}}}\right)^{\Delta_\phi} \sum_{m=0}^{\infty} (1-x)^m \sum_{k=-m}^{m} B^{(m,k)}_{\Delta,\ell} \frac{\kappa_{\Delta+\ell+2k}}{\kappa_{\Delta+\ell}} k_{\Delta+\ell+2k}(1-\bar{x})\,.
\end{aligned}\end{equation}
They can either be calculated recursively using the quadratic Casimir equation like in \cite{Caron-Huot:2017vep} or through
the series expansion \eqref{eq:HSfct}. The first few coefficients are
\begin{equation}\label{eq:Bcoeff}\begin{aligned}
B_{\Delta,\ell}^{(0,0)} &= 1 \,,\\
B_{\Delta,\ell}^{(1,-1)} &= \frac{(\Delta -3) (d-2 p-2) (\Delta +l-2)^2}{16 (d-2 \Delta +2) (\Delta +\ell-3) (\Delta +\ell-1)} \,,\\
B_{\Delta,\ell}^{(1,0)} &= \frac{\ell-\Delta +2}{4} \,,\\
B_{\Delta,\ell}^{(1,1)} &= -\frac{(\ell+2) (d-2 p-2)}{d+2 \ell} \,.
\end{aligned}\end{equation}
Our goal is to compute the leading contribution to the partial wave amplitude $c(\Delta,\ell)$ in
the lightcone limit, under the assumptions on the defect theory we formulated and discussed at the
beginning of this section. In more mathematical terms, these assumptions imply that the defect
channel partial wave decomposition of the function $\mathcal{F}$ has the form
\begin{equation} \label{eq:defidentityAnsatz}
 \mathcal{F}(x,\bar{x}) = 1 + \ \sum_{\widehat{\Delta}-s \geq \delta} b_{ \widehat{\mathcal{O}}}^2
 \hat{f}_{\widehat{\Delta},s}(x, \bar{x}) \,,
\end{equation}
where the sum extends over defect primaries $\widehat{\mathcal{O}}$ whose weight $\widehat{\Delta}$
and spin $s$ satisfy $\widehat{\Delta}-s \geq \delta$ for some positive finite real number $\delta$. If
$\mathcal{F}$ has this form, only the first constant terms actually contributes to the leading
lightcone behavior. Let us first determine this contribution $C^t_0$ to the generating function
\eqref{eq:genFctDef},
\begin{equation}\label{eq:shdowPolesDef}\begin{aligned}
C^t_0(x,\beta) &= (1-x)^{\Delta_\phi} \kappa_\beta \int_{0}^{1}\! d\bar{x}\,  \bar{x}^{-\frac{\Delta_\phi}{2}}(1-\bar{x})^{\Delta_\phi-2} k_\beta(1-\bar{x}) \operatorname{dDisc} 1 \\
&= (1-x)^{\Delta_\phi}  \frac{\sin^2\left(\pi\frac{\Delta_\phi}{2}\right)}{\pi^2}
\frac{\Gamma^4\left( \frac{\beta}{2} \right)}{\Gamma\left( \beta-1 \right)\Gamma\left( \beta \right)}\frac{\Gamma\left(1-\frac{\Delta_\phi}{2}\right)\Gamma\left(\frac{\beta}{2}+\Delta_\phi-1\right)}{\Gamma\left(\frac{\beta}{2}+\frac{\Delta_\phi}{2}\right)}
\\
&\qquad\times
\pFq{3}{2}{\frac{\beta}{2},\frac{\beta}{2},\frac{\beta}{2}+\Delta_\phi-1}{\beta,\frac{\beta}{2}+\frac{\Delta_\phi}{2}}{1} \\
&= \frac{(1-x)^{\Delta_\phi}}{2^{\frac{\beta}{2}-\Delta_\phi+1} \Gamma^2\left( \frac{\Delta_\phi}{2} \right) } \frac{\Gamma^2\left( \frac{\beta}{4} \right) }{ \Gamma\left( \frac{\beta}{2}-\frac12 \right)} \frac{ \Gamma\left( \frac{\beta}{4}+\frac{\Delta_\phi}{2} -\frac12 \right)}{\Gamma\left( \frac{\beta}{4}-\frac{\Delta_\phi}{2} +1 \right)} \,,
\end{aligned}\end{equation}
where we used  Watson's theorem to write the generalized hypergeometric function ${}_3F_2$ in terms of gamma functions \cite{Bateman:1981}.
We see that equation \eqref{eq:shdowPolesDef} has poles in $1-x$ whenever $\Delta-\ell=2\Delta_\phi$. Like in the four-point case, the family of double-twist operators emerge in the bulk spectrum at large spin. They take the schematic form
\begin{equation}
[\phi^2]_{n,\ell} \sim \phi (\partial^2)^n \partial_{\mu_1} \dots \partial_{\mu_\ell} \phi \,.
\end{equation}
Including the identical $u$-channel contribution, equation \eqref{eq:resCDef} yields for the product of operator
product coefficients
\begin{equation}
a_{[\phi^2]_{0,\ell}} c_{\phi\phi[\phi^2]_{0,\ell}} = \frac{(1+(-1)^\ell)}{2^{\ell+1} \left(\frac{\ell}{2}\right)!} \frac{\left(\frac{\Delta_\phi}{2}\right)_\frac{\ell}{2}^2}{\left(\Delta_\phi+\frac{\ell}{2}-\frac12\right)_\frac{\ell}{2}}\,,\qquad \ell \rightarrow \infty \,,
\end{equation}
where $c_{\phi\phi[\phi^2]_{0,\ell}}$ is given by the mean field coefficients \cite{Fitzpatrick:2012yx}. Since the latter
are only determined up to a sign, so are the one-point coefficients $a_{[\phi^2]_{0,\ell}}$. The final result is
\begin{equation}
\left( a_{[\phi^2]_{0,\ell}} \right)^2 = \frac{(1+(-1)^\ell)}{8} \frac{\ell!}{\left[\left(\frac{\ell}{2}\right)!\right]^2} \frac{\left( \frac{\Delta_\phi}{2} \right)_\frac{\ell}{2}^4}{\left( \Delta_\phi \right)_\frac{\ell}{2}^2} \frac{1}{(2\Delta_\phi+\ell-1)_l} \,,\qquad \ell \rightarrow \infty \,.
\end{equation}
Let us stress that this behavior is quite general provided our assumptions are satisfied: they do not
depend on the details of the defect theory. Furthermore, the analyticity in spin shows that every
operator in the double-twist family $[\phi^2]_{0,\ell}$ couples to the defect.
\medskip

One would be tempted to continue this kind of analysis and determine lower-order corrections to the
leading lightcone behavior by considering more defect blocks with higher twist in the decomposition in equation
\eqref{eq:defidentityAnsatz}. However, it turns out that most features of the bulk cannot be obtained
from a single defect block but require an infinite set of such blocks to conspire. In particular, a single
defect block \eqref{eq:defBlock} does not have a logarithmic behavior in the limit $x \rightarrow 1$. On
the other hand, such logarithms must appear whenever (high spin) bulk fields possess anomalous contributions
to their conformal weights, as it is the case for most operators in interacting theories. In order to account for
such logarithms, we need an infinite family of defect operators with approximately same twist, and with
\eg bulk-to-defect coefficients going like $b^2_s \rightarrow 1/s$ in the limit $s \rightarrow \infty$,
\begin{align}
	\sum_{s}^\infty b_s^2 \hat{f}_{\widehat{\Delta}+s,s}(x,\bar{x}) &\stackrel{\bar{x}\rightarrow 0}
{\rightarrow} \sum_{s}^\infty b_s^2 x^{\frac{\widehat{\Delta}+s}{2}}
\bar{x}^{\frac{\widehat{\Delta}-s}{2}} \\[2mm]
	&\stackrel{x\rightarrow 1}{\rightarrow}  - b_\odot^2 \bar{x}^{\frac{\widehat\tau}{2}} \log(1-x)
\,,\qquad\text{for }b_s^2 \approx \frac{b_\odot^2}{s} \,,
\end{align}
where we evaluated the defect blocks $\hat{f}$ in the double-light cone limit $(x,\bar{x})\sim(1,0)$.
Therefore we have to sum over the family first before inverting. A similiar issue arises in
the context of two-point functions at finite temperature \cite{Iliesiu:2018fao}, we leave a more
systematic study of such infinite sums of defect blocks to future work.  

%% file: 3dIsing.tex
%!TEX root = ../bulk_inversion.tex
%%%%%%%%%%%%%%%%%%%%%%%%%%%%%%%%%%%%%%%%%%%%%

\section{$3d$ Ising twist defect}
\label{sec:3dIsing_twist_defect}

In this section we apply our inversion formula to the twist defect of the $3d$ Ising model. The 
twist defect can be constructed in the lattice \cite{Billo:2013jda} as the boundary of a surface 
of frustrated links or, in a dual description, the Wilson line of the $\mathbb{Z}_2$ gauge theory.
The continuum limit is expected to be described by a defect CFT that was further 
studied in \cite{Gaiotto:2013nva} through the $\epsilon$-expansion of Wilson and Fisher as well as 
the numerical bootstrap. Here we will determine the full dynamical content in the bulk channel for 
a two-point function of spin fields $\phi = \sigma$ in the presence of the twist defect, to leading 
nontrivial order in $\epsilon = 4-d$. More concretely, we determine analytically the bulk-to-defect 
couplings of all the bulk primaries that can appear in the OPE of two spin fields to 
linear order in $\epsilon$.  

Let us stress that the dynamical content of the bulk expansion $\sigma \times \sigma = 1 + \varepsilon
+ \dots$ of the spin field $\sigma$ has been studied by a variety of techniques, and in particular the 
numerical bootstrap has given the most precise estimates of the conformal dimension of the spin field 
$\sigma$, the energy operator $\varepsilon$ and other fields along with their OPE coefficients \cite{ElShowk:2012ht,El-Showk:2014dwa,Kos:2014bka,Simmons-Duffin:2015qma,Kos:2016ysd}. On the other hand, 
much less is known about OPEs or block decompositions in the presence of the twist defect.

The starting point of our analysis is a scalar field in $d=4-\epsilon$ dimensions in the presence of 
a twist defect of dimension $p=2-\epsilon$. By definition, the spin field $\phi = \sigma$ acquires a sign 
when transported around the twist defect. If the system is now perturbed by the relevant operator $\phi^4$
one expects the infrared theory to belong to the same universality class as the twist defect of the discrete
Ising model. Because the defect does not change the renormalization group flow locally, the bulk theory has 
the standard coupling $g = (4\pi)^3\epsilon/3 +  \mathcal{O}(\epsilon^2)$; this analysis was performed 
to one-loop order in \cite{Gaiotto:2013nva}. Below we will review the existing results from the 
$\epsilon$-expansion, with some small improvements that will simplify the subsequent analysis.

\subsection{Free theory two-point functions}

The two-point function of the fundamental field $\phi$, in the present context often referred to as the 
spin operator, in a $d$-dimensional free scalar field theory is given by
\beq
\label{eq:exact_two_pt_function}
\begin{aligned}
	\mathcal{F}^\text{free}_\text{twist}(x,\bar{x}) &=	(r_1r_2)^{\Delta_\phi}\langle \mathcal{D}^{(d-2)}_{\text{twist}} \phi(x_1) \phi(x_2) \rangle \\
	&= \left( \frac{(1-x)(1-\bar{x})}{\sqrt{ x\bar{x}}} \right)^{-\Delta_\phi} \operatorname{I}\left( \left( \frac{ \sqrt x + \sqrt{\bar{x}} }{ 1+\sqrt{x\bar{x}} } \right)^2; \frac12, \Delta_\phi \right)\, ,
\end{aligned}
\eeq
where we abbreviated $r_i = |x_i^\perp|$ and $I(x;a,b)=I_x(a,b)$ is the regularized incomplete beta function. When the dimension of the scalar field $\phi$ is fixed to its free 
field theory value, i.e.\ we set  $\Delta_{\phi} = \frac{d}{2}-1$, the conformal block expansion reads
\begin{equation}\begin{aligned}
\mathcal{F}^\text{free}_\text{twist}(x,\bar{x}) &=\sum_{s=\frac12}^{\infty} b^2_s  \hat{f}_{\Delta_\phi+s,s}(x,\bar{x})\,, \\[2mm] 
&= \left( \frac{(1-x)(1-\bar{x})}{\sqrt{ x\bar{x}}} \right)^{-\Delta_\phi} \left( 1 + \sum_{\ell=0,2,\dots} 
\alpha^{\text{free}}_{\ell} f_{2\Delta_\phi+\ell,\ell}(x,\bar{x}) \right) \,, 
\end{aligned}\end{equation}
where $\alpha^{\text{free}}_{\ell} = a_{[\phi^2]_{0,\ell}} c_{\phi\phi[\phi^2]_{0,\ell}}$.
Here, the first line is the defect channel expansion \eqref{eq:defectCBD} while the second line performs a partial 
wave decomposition \eqref{eq:2ptbulkCBD} in the bulk channel. The relevant defect OPE coefficients 
are given by \cite{Gaiotto:2013nva}
\begin{align}
b^2_s &= \frac{(\Delta_\phi)_s}{s!} \,,
\end{align}
whereas we used our inversion formula to obtain the bulk OPE coefficients for $\ell > 1$
\begin{align}
\alpha^{\text{free}}_\ell &= \frac{1+(-1)^\ell}{2^{2\Delta_\phi +2\ell+2}\pi} \frac{\Gamma \left(\frac{\ell}{2}-\frac{1}{2}\right)}{\left(\frac{\ell}{2}\right)! } \frac{\Gamma \left(\Delta_\phi +\frac{1}{2}\right)  }{\Gamma \left(\Delta_\phi \right)  } \frac{\left(\Delta_\phi +\frac{\ell}{2}+1\right)_{\frac{\ell}{2}-1}}{\left(\Delta_\phi +\frac{\ell}{2}-\frac{1}{2}\right)_\frac{\ell}{2}} \,. 
\end{align}
The inversion formula does not capture the case $\ell = 0$. However, analytic continuation shows that the coefficient $\alpha^{\text{free}}_0$ matches the one calculated in \cite{Gaiotto:2013nva}. We should point put out that these results already represent a small improvement of the formulas in \cite{Gaiotto:2013nva}, where the resummed propagator was written only for the cases $d=4$ and $x = \bar{x}$. 

\paragraph{An exact solution to crossing.}
Before we proceed with the $\epsilon$-expansion analysis, let us point out an interesting consequence of the exact formula \eqref{eq:exact_two_pt_function}. Instead of fixing $\Delta_{\phi}$ to its free-field value, we can consider an analytic continuation for arbitrary values of $\Delta_{\phi}$. This exact formula has a sensible expansion in conformal blocks in both bulk and defect channels (see below), and can therefore be thought of as a defect two-point function. We call this solution \textit{mean-field twist defect}. This formula not only generalizes the free-theory defect, but it is also a natural counterpart to the \textit{trivial defect} correlator which consists of only the identity operator in the bulk:
\begin{align}
\label{eq:trivial_defect}
\mathcal{F}_\text{triv}(x,\bar{x}) = (r_1r_2)^{\Delta_\phi}\langle \mathbf{I} \phi(x_1) \phi(x_2) \rangle &= \left( \frac{(1-x)(1-\bar{x})}{\sqrt{ x\bar{x}}} \right)^{-\Delta_\phi}\, .
\end{align} 
A way to see the relation between these solutions is by looking at the defect channel expansion:
\begin{align}
\mathcal{F}^\text{mf}_\text{twist}(x,\bar{x}) &=  \sum_{m=0}^\infty \sum_{s=\frac12}^{\infty} b^2_{m,s}  \hat{f}_{\Delta_\phi+2m+s,s}(x,\bar{x})\, ,
\\
\mathcal{F}_\text{triv}(x,\bar{x}) & =  \sum_{m=0}^\infty \sum_{s=0}^{\infty} b^2_{m,s}  \hat{f}_{\Delta_\phi+2m+s,s}(x,\bar{x})\, .
\end{align}
Here the $b^2_{m,s}$ coefficients are identical and given by
\begin{equation}
b^2_{s,m}=\frac{ \left(\D_\phi-\frac{d}{2}+1\right)_m (\D_\phi)_{2 m+s}
	}{ m! s!
	\left(s+\frac{d-p}{2}\right)_m  \left( \D_\phi + m+s-\frac{p}{2}\right)_m}\,.
\label{btrivial}
\end{equation}
The only difference between the two expansions is the range of the spin variable: integer spins for the trivial defect and half-integer spins for the mean-field twist defect. The bulk channel expansions are of course quite different, the trivial defect has only one term (by definition),
 while the mean-field twist defect has an infinite number of terms:
\begin{align}
\mathcal{F}^\text{mf}_\text{twist}(x,\bar{x})  & = \left( \frac{(1-x)(1-\bar{x})}{\sqrt{ x\bar{x}}} \right)^{-\Delta_\phi} \left( 1 + \sum_{n=0}^\infty \sum_{\ell=0,2,\dots} \alpha^{\text{mf}}_{n,\ell} f_{2\Delta_\phi+2n+\ell,\ell}(x,\bar{x}) \right) \,.
\end{align}
We do not have a close-form expression for the $\alpha^{\text{mf}}_{n,\ell}$ 
but we state the first few of them in appendix \ref{apx:mft_coeff}.

\subsection{One-loop correction and bulk inversion}

The one-loop correction to the two-point function was obtained in integral form in \cite{Gaiotto:2013nva}. 
Here we improve on that result by presenting a closed-form expression for this correlator,
\beq
\label{eq:one_loop_correlator}
\mathcal{F}^\text{1-loop}_\text{twist}(x,\bar{x}) =  - \frac{\epsilon}{24} \frac{\sqrt{x\bar{x}}}{1-x\bar{x}} \log(x\bar{x}) \operatorname{arctanh}\left( \frac{ \sqrt x + \sqrt{\bar{x}} }{ 1+\sqrt{x\bar{x}} } \right) \,.
\eeq
Let us make some comments on the derivation of this formula. Using the \textit{defect channel} inversion 
formula of \cite{Lemos:2017vnx}, one can obtain defect data in terms of a (single) discontinuity of the 
correlator. At order $\epsilon$, only $\phi^2$ has an anomalous dimension and therefore the discontinuity 
operation $\operatorname{Disc}$ in the defect channel inversion formula kills all other possible 
contributions. The $\phi^2$ term gives the leading twist family $\hat{\Delta}=\Delta_{\phi} + s$, and 
one can show by explicit calculation that higher-twist families on the defect are of order 
$\mathcal{O}(\epsilon^2)$. This fact implies that the leading twist family is all there is at leading 
order. Now, for low enough values of the transverse spin variable $s$, the analytic structure of the 
CFT data might change, because the validity of the inversion formula of 
\cite{Lemos:2017vnx} is tied to the behavior of the correlator in the limit $w \to 0$, which is the 
analog of the Regge limit of four-point functions in bulk CFTs \cite{Costa:2012cb,
Caron-Huot:2017vep}.\footnote{Note that the definition of $w$ in \cite{Lemos:2017vnx} differs from the one we used in section \ref{sec:derivation}} In \cite{Lemos:2017vnx}, by numerically integrating the one-loop correction of 
\cite{Gaiotto:2013nva}, it was argued that the correlator behaves as $w^{0}$ in the $w \to 0$ limit, 
which then implies that the analytic defect data obtained from the defect channel inversion formula 
is valid for $s>0$. Consequently, the leading-twist family $\hat{\Delta}=\Delta_{\phi} + s$ describes 
the \textit{full} two-point function at one-loop, with no additional contributions. It is then an 
easy exercise to resum the defect channel block expansion to obtain equation \eqref{eq:one_loop_correlator}.

Armed with the explicit one-loop expression we will now use the \textit{bulk channel} inversion 
formula \eqref{eq:defInversion} to obtain analytic expression for the bulk channel data. Evaluating 
the relevant discontinuity operation $\operatorname{dDisc}$ of the one-loop formula \eqref{eq:one_loop_correlator} 
gives\footnote{We used the identity $\op{arctanh}(\sqrt x)+\op{arctanh}(\sqrt{\bar{x}}) = \operatorname{arctanh}\left( \frac{ \sqrt x + \sqrt{\bar{x}} }{ 1+\sqrt{x\bar{x}} } \right)$.} 
\begin{equation}
\operatorname{dDisc} \mathcal{F}^\text{1-loop}_\text{twist}(x,\bar{x}) = -\frac{\epsilon}{12} \frac{\sqrt{x\bar{x}}}{1-x\bar{x}} \log(x\bar{x}) \operatorname{arctanh}\left( \sqrt{\bar{x}} \right)\,.
\end{equation}
Since the discontinuity operation $\operatorname{dDisc}$ kills the term $$\operatorname{arctanh}\left( 
\sqrt{x} \right) \stackrel{x \rightarrow 1}{\rightarrow} -\log(1-x) \,, $$ there is no anomalous dimension for higher-spin operators on the bulk side, as expected. 
The bulk OPE coefficients can be directly calculated using equations \eqref{eq:genFctDef} and 
\eqref{eq:genFctExp}. In the defect channel the conformal block expansion reads
\begin{equation}
\begin{aligned}
& \mathcal{F}_\text{twist}(x,\bar{x}) = \mathcal{F}^\text{free}_\text{twist}(x,\bar{x}) + \mathcal{F}^\text{1-loop}_\text{twist}(x,\bar{x}) = \sum_{s=\frac12}^{\infty} b^2_s  \hat{f}_{ \Delta_s ,s}(x,\bar{x}) + \dots\\[2mm] 
& \quad \quad = \left( \frac{(1-x)(1-\bar{x})}{\sqrt{ x\bar{x}}} \right)^{-\Delta_\phi} \left( 1 + \underbrace{ \sum_{\ell=0,2,\dots}  \alpha_{\ell} f_{\Delta^{(2)}_\ell,\ell}(x,\bar{x}) }_{[\phi^2]_{0,\ell}} + \underbrace{ \sum_{\ell=0,2,\dots} \beta_{\ell} f_{\Delta^{(4)}_\ell,\ell}(x,\bar{x}) }_{[\phi^4]_{0,\ell}} \right) + \dots \, , 
\end{aligned}
\end{equation}
up to terms of order $\mathcal{O}(\epsilon^2)$. Here, $\Delta_\phi = 1 - \frac{\epsilon}{2}$ and the 
defect data is \cite{Gaiotto:2013nva}
\begin{align}
\hat{\Delta}_s &= 1 + s - \left(\frac12 + \frac{1}{24s}\right)\epsilon \,,\\
b^2_s &= \frac{(1-\frac{\epsilon}{2})_s}{s!} \,.
\end{align}
The bulk conformal dimensions are well-known,
\begin{align}
	\Delta^{(2)}_\ell &= \begin{cases}
	2 - \frac{2}{3}\epsilon & \text{if $\ell = 0$} \,,\\
	2 - \epsilon + \ell & \text{if $\ell \geq 2$} \,,\\
	\end{cases}	\\
	\Delta^{(4)}_\ell &= 4 - 2\epsilon + \ell \,.
\end{align}
New are the OPE coefficients obtained by the inversion formula (except for $\alpha_0$ which is given in \cite{Gaiotto:2013nva}),
\begin{align}
\frac{\alpha_\ell}{\alpha_\ell^{\text{free}}} &= 1+ \epsilon\begin{cases}
	-\frac{2}{3} \log 2 & \text{if $\ell = 0$} \,,\\
	\frac{1}{384}\frac{(\ell-1) (\ell+2)}{\ell (\ell+1)} \Gamma^3 \left(\frac{\ell}{2}+1\right) \Gamma \left(\frac{\ell}{2}\right) (-f_1(\ell)-f_2(\ell)+3f_3(\ell)+2f_4(\ell)) & \text{if $\ell \geq 2$} \,,
\end{cases}\\
\beta_\ell & = \epsilon\frac{ 2^{-(6+\ell)} }{ 3\pi (\ell+2)^2 } \frac{ \Gamma\left(\frac\ell2+\frac12\right)^2 }{ \left( \frac\ell2 \right)! } \frac{ \Gamma\left( \frac\ell2+\frac32 \right) }{ \Gamma\left( \ell+\frac32 \right) } \,.
\end{align}
where
\begin{equation}\label{eq:1loopHyper}
		f_k(\ell) = \frac{\pi ^2 \, \pFq{4}{3}{\overbrace{\scriptstyle \frac12,\dots,\frac12}^{4-k},\overbrace{\scriptstyle \frac32,\dots,\frac32}^{k}}{\scriptstyle 2,2,2}{1^-}}{\Gamma \left(\frac{\ell}{2}+\frac{1}{2}\right)^3 \Gamma \left(\frac{\ell}{2}+\frac{3}{2}\right)}
		-\frac{2 (\ell+1)^{k-1} \, \pFq{5}{4}{\scriptstyle 1,\overbrace{\scriptstyle \frac{\ell}{2}+\frac{1}{2},\dots,\frac{\ell}{2}+\frac{1}{2}}^{4-k} \overbrace{\scriptstyle \frac{\ell}{2}+\frac{3}{2},\dots,\frac{\ell}{2}+\frac{3}{2}}^{k}}{\scriptstyle \frac{\ell}{2}+1,\frac{\ell}{2}+2,\frac{\ell}{2}+2,\frac{\ell}{2}+2}{1^-} }{ \Gamma \left(\frac{\ell}{2}+1\right) \Gamma^3 \left(\frac{\ell}{2}+2\right) }\,.
\end{equation}
The individual hypergeometric functions may diverge at unity argument but their combination is finite in equation \eqref{eq:1loopHyper}. Using the known three-point coefficients (see \eg \cite{Alday:2017zzv}), one can solve for the one-point functions $a_{[\phi^2]_{0,\ell}}$ and $a_{[\phi^4]_{0,\ell}}$ at order $\mathcal{O}(\epsilon)$ and $\mathcal{O}(1)$, respectively. This concludes our analysis of the twist defect in the $3d$ Ising model.

%% file: conclusions.tex
%!TEX root = ../bulk_inversion.tex
%%%%%%%%%%%%%%%%%%%%%%%%%%%%%%%%%%%%%%%%%%%%%

\section{Conclusions and outlook}
\strut

In this work we have derived a Lorentzian inversion formula for the bulk channel of defect CFTs. The formula captures the bulk CFT data, which consists of products of three-point coupling and one-point functions, as an analytic function of the spin variable $\ell$. Similar to four-point functions in bulk CFTs, the defect data is obtained as the poles and residues of an integral of a certain discontinuity of the correlator. Because three-point couplings were already known to be analytic in spin,
our result demonstrates that one-point functions also sit in analytic trajectories of the spin variable.

Our bulk channel inversion formula can now be used with the defect channel inversion formula of \cite{Lemos:2017vnx} to implement the bootstrap program for defect CFTs, starting from one channel and the alternating between the two formulas.
We already saw a first application of this procedure in \mbox{section \ref{sec:3dIsing_twist_defect}}, where we bootstrapped (under mild assumptions) the one-loop correlator for the Ising model twist. Two natural follow-ups would be to bootstrap higher orders corrections and also generalize the result to $O(N)$ models, similar to what was done for four-point functions in \cite{Alday:2017zzv,Henriksson:2018myn} (see also \cite{Gopakumar:2016wkt,Gopakumar:2016cpb,Dey:2016mcs} for the Mellin approach to the same problem). This procedure can also be applied to other defect setups of interest. One possible application is the Yang-Lee model which is amenable to $\epsilon$-expansion techniques and has already been studied using modern bootstrap techniques \cite{Gliozzi:2014jsa}. Finally, another interesting case are line defects in $\mathcal{N}=4$ super Yang-Mills, where the relevant superblocks were calculated in \cite{Liendo:2016ymz}.

Another generalization of this work is to consider external operators with spin.
Important examples of spinning operators are conserved flavor currents and the stress-energy
tensor. The understanding of spinning blocks for bulk four point functions has advanced 
significantly since the early papers on the subject \cite{Costa:2011dw,Costa:2011mg,
SimmonsDuffin:2012uy,Costa:2014rya}. Subsequent developments include the concept and 
construction of seed blocks \cite{Echeverri:2015rwa,Echeverri:2016dun,Costa:2016hju,
Cuomo:2017wme}, the Calogero-Sutherland approach to spinning blocks \cite{Schomerus:2016epl,
Schomerus:2017eny} as well as the introduction of weight shifting operators in 
\cite{Karateev:2017jgd}. A Lorentzian inversion formula for spinning four-point correlators 
has also been obtained in \cite{Kravchuk:2018htv} through the investigation of light-ray 
operators. Using input from the Calogero-Sutherland approach to spinning four-point blocks
\cite{Schomerus:2016epl,Schomerus:2017eny} it is clear how to extend the setup of
\cite{Isachenkov:2018pef} to develop a systematic theory of spinning blocks for
defect two-point functions, both in the bulk and the defect channel, extending a
recent studies in \cite{Lauria:2018klo}. Following the approach we have developed
above, it should then be possible to obtain a Lorentzian inversion formulas for
spinning defect two-point functions that is consistent with the Lorenztian 
inversion formula for spinning four-point functions from \cite{Kravchuk:2018htv}. It would be 
interesting to work out the details and apply such a Lorentzian inversion 
formula, in particular to bootstrapping two-point functions of the stress 
tensor in the presence of a defect.

\acknowledgments
We wish to thank Ilija Buric, Misha Isachenkov, 
Madalena Lemos, Marco Meineri and Evgeny Sobko for comments and discussion. The work of PL is supported by the DFG through the Emmy Noether research group ``The Conformal Bootstrap Program''.

%% file: AppendixA.tex
\section{Bulk OPE coefficients of the mean field twist defect}
\label{apx:mft_coeff}

In this appendix we want to outline the calculation of the bulk OPE coefficients $\alpha^{\text{mf}}_{n,\ell}$ for the mean-field twist defect two-point function \eqref{eq:exact_two_pt_function}.
The coefficient $\alpha^{\text{mf}}_{0,\ell}$ is given by its free-field counterpart since the two-point function does not depend on the dimension $d$,
\begin{equation}
	\alpha^{\text{mf}}_{0,\ell} = \alpha^{\text{free}}_{0,\ell} \,.
\end{equation}
The mean-field coefficients beyond $n=0$ follow from our inversion formula by including sub-leading powers of $(1-x)$. However, instead of attacking the relevant integrals directly, we can save some effort by using again that the two-point function does not depend on the dimension $d$ --- its dependence enters through the coefficients $B^{(m,k)}_{\Delta,\ell}$ in equation \eqref{eq:Bcoeff}. Hence, we can solve for the integrals by demanding that the OPE coefficients $\alpha_{n,\ell}$ vanish for $n>0$ in free field theory, \ie setting $d=2\Delta_\phi+2$ in the coefficients $B^{(m,k)}_{\Delta,\ell}$. Once the integrals are obtained, we can calculate the $\alpha_{n,\ell}$ by leaving $d$ arbitrary. The first two mean-field coefficients beyond $n=0$ are
\begin{equation}\begin{aligned}
\alpha_{n,\ell}^{\text{mf}} &=\frac
{\left[1+(-1)^\ell\right] \left(\Delta_\phi-\frac{d}{2} +1\right)_n \left(\Delta_\phi\right)_{n+\ell} \Gamma \left(\frac{\ell-1}{2}\right)  }
{\left(\frac{\ell}{2}\right)! \left(\ell+\frac{d}{2}\right)_n \left( 2\Delta_\phi +\ell+n-\frac{d}{2} \right)_n \left( \Delta_\phi + n+\frac{\ell}{2} - \frac{1}{2} \right)_{\frac{\ell}{2}} (\Delta_\phi+\frac12)_{n+\frac{\ell}{2}+\frac12}} \\
&\qquad\times
\frac{p_{n,\ell}}{ 2^{2 \Delta_\phi +2 \ell+4 n+2} \pi} \,,
\end{aligned}\end{equation}
where
\begin{align}
&\begin{aligned}
	p_{1,\ell} &= -2\Delta_\phi(d+2\ell)-2\ell(\ell+1)-d+2) \,,
\end{aligned}\\
&\begin{aligned}
	p_{2,\ell} &= \frac{1}{-4 \Delta_\phi+2d -6} \Big( 4 d^3 \Delta_\phi ^2+8 d^3 \Delta_\phi +3 d^3-8 d^2 \Delta_\phi ^3+2 d^2 \Delta_\phi  \left(4 \ell^2+16 \ell-11\right)\\
	&\qquad +4 d^2 \Delta_\phi ^2 (4 \ell-5)+d^2 (2 \ell-1) (2 \ell+7)+4 d \Delta_\phi  \left(4 \ell^3+6 \ell^2-26 \ell-13\right) \\
	&\qquad +2 d \left(2 \ell^4+12 \ell^3+10 \ell^2-24 \ell-15\right)-16 d \Delta_\phi ^3 (2 \ell+1)-8 d \Delta_\phi ^2 (12 \ell+5) \\
	&\qquad -16 \Delta_\phi ^2 (\ell+1) \left(2 \ell^2+7 \ell-2\right)-4 (\ell+1) (\ell+2) \left(2 \ell^2+6 \ell-5\right) \\
	&\qquad -8 \Delta_\phi  (\ell+1) \left(\ell^3+9 \ell^2+14 \ell-9\right)-32 \Delta_\phi ^3 \ell (\ell+1) \Big) \,.
\end{aligned}
\end{align}